\documentclass[aps,preprint,superscriptaddress,12pt]{revtex4-2}

\usepackage{graphicx}
\usepackage{amsmath}
\usepackage{amssymb}
\usepackage[hidelinks]{hyperref}
\usepackage{xcolor}
\usepackage[normalem]{ulem}
\usepackage{float}

\begin{document}


\title{Symmetry breaking of large-amplitude parametric oscillations in few-layer graphene nanomechanical resonators}

\author{Chen Yang}
\affiliation{School of Optoelectronic Science and Engineering \& Collaborative Innovation Center of Suzhou Nano Science and Technology, Soochow University, 215006 Suzhou, China}
\affiliation{Key Lab of Advanced Optical Manufacturing Technologies of Jiangsu Province \& Key Lab of Modern Optical Technologies of Education Ministry of China, Soochow University, 215006 Suzhou, China}

\author{YuBin Zhang}
\affiliation{School of Optoelectronic Science and Engineering \& Collaborative Innovation Center of Suzhou Nano Science and Technology, Soochow University, 215006 Suzhou, China}
\affiliation{Key Lab of Advanced Optical Manufacturing Technologies of Jiangsu Province \& Key Lab of Modern Optical Technologies of Education Ministry of China, Soochow University, 215006 Suzhou, China}

\author{Heng Lu}
\affiliation{School of Optoelectronic Science and Engineering \& Collaborative Innovation Center of Suzhou Nano Science and Technology, Soochow University, 215006 Suzhou, China}
\affiliation{Key Lab of Advanced Optical Manufacturing Technologies of Jiangsu Province \& Key Lab of Modern Optical Technologies of Education Ministry of China, Soochow University, 215006 Suzhou, China}

\author{Ce Zhang}
\affiliation{School of Optoelectronic Science and Engineering \& Collaborative Innovation Center of Suzhou Nano Science and Technology, Soochow University, 215006 Suzhou, China}
\affiliation{Key Lab of Advanced Optical Manufacturing Technologies of Jiangsu Province \& Key Lab of Modern Optical Technologies of Education Ministry of China, Soochow University, 215006 Suzhou, China}

\author{FengNan Chen}
\affiliation{School of Optoelectronic Science and Engineering \& Collaborative Innovation Center of Suzhou Nano Science and Technology, Soochow University, 215006 Suzhou, China}
\affiliation{Key Lab of Advanced Optical Manufacturing Technologies of Jiangsu Province \& Key Lab of Modern Optical Technologies of Education Ministry of China, Soochow University, 215006 Suzhou, China}

\author{Ying Yan}
\affiliation{School of Optoelectronic Science and Engineering \& Collaborative Innovation Center of Suzhou Nano Science and Technology, Soochow University, 215006 Suzhou, China}
\affiliation{Key Lab of Advanced Optical Manufacturing Technologies of Jiangsu Province \& Key Lab of Modern Optical Technologies of Education Ministry of China, Soochow University, 215006 Suzhou, China}

\author{Fei Xue}
\affiliation{School of Physics, Hefei University of Technology, 230601 Hefei, China}

\author{Alexander Eichler}
\affiliation{Laboratory for Solid State Physics, ETH Z\"{u}rich, 8093 Z\"{u}rich, Switzerland}

\author{Joel Moser}
\email{j.moser@suda.edu.cn}
\affiliation{School of Optoelectronic Science and Engineering \& Collaborative Innovation Center of Suzhou Nano Science and Technology, Soochow University, 215006 Suzhou, China}
\affiliation{Key Lab of Advanced Optical Manufacturing Technologies of Jiangsu Province \& Key Lab of Modern Optical Technologies of Education Ministry of China, Soochow University, 215006 Suzhou, China}




\date{\today}

\begin{abstract}
Graphene nanomechanical resonators are well suited for the study of parametric oscillations. Their large frequency tunability and their pronounced nonlinearities enable an efficient modulation of their resonant frequencies. Here, we present measurements of the response of few-layer graphene nanomechanical resonators, each driven by a large parametric pump at frequency $2\omega$ and a weak external drive at $\omega$, where $\omega$ is set near the mechanical resonant frequency $\omega_0$. The pump actuates the resonator beyond the threshold for large-amplitude parametric oscillations, while the drive breaks the symmetry between the parametric phase states. By increasing and decreasing a gate voltage to detune $\omega_0$ in the presence of the pump and the drive, we observe a double hysteresis in the response. The double hysteresis reveals the existence of two possible large-amplitude vibrational states whose phase difference is nearly $\pi$ radians. We deterministically prepare the resonator in either one of these states by cycling the gate voltage. We measure the stationary occupation probabilities of the two states in the presence of a white Gaussian force noise, and find that they strongly depend on the amplitude and on the phase of the external drive. The phase states of parametric oscillations with broken amplitude symmetry can be mapped to biased bi-modal degrees of freedom, such as Ising spins in an external magnetic field. Therefore, they hold promise as units of binary information.
\end{abstract}


\maketitle

A mechanical parametric resonance is the state of a resonator driven by the modulation of its spring constant beyond a threshold \cite{Rugar1991,Turner1998,Carr2000,Karabalin2010,Eichler2012,Gieseler2012}. The dynamics of a parametric resonator results from an interplay between the parametric excitation (the pump), the linear and the nonlinear restoring forces, and mechanical dissipation \cite{Lifshitz2009,Eichler2023}. It is governed by the depth $\lambda$ and the angular frequency $\omega_\mathrm{p}$ of the modulation of the spring constant $k_0$ as a function of time $t$, $k_0\rightarrow k_0(1-\lambda\cos\omega_\mathrm{p}t)$. Where $\lambda$ is large enough that the rate of energy supplied by the pump compensates for dissipation, and with $\omega_\mathrm{p}$ close to twice the vibrational resonant frequency $\omega_0$, the effective damping of the resonator becomes negative and the resonator enters a regime of large-amplitude oscillations near $\omega_0$ \cite{Lifshitz2009}. In the absence of an external drive near $\omega_0$, the parametric resonator hosts one of two possible vibrational states whose amplitudes are equal and whose phases differ by $\pi$ radians. Both states are equally likely to be actuated by the pump --experimentally, this can be seen by injecting a force noise to toggle the resonator between its two stable states stochastically, resulting in two equally distributed populations \cite{Dykman1998,Lapidus1999,Marthaler2006,Kim2006,Chan2007,Mahboob2010b,Han2024}. These `phase states' are interesting, in part because they are the states of the parametron employed in early digital computing systems \cite{Goto1959}. There, binary information was encoded in the phase of electrical parametric oscillations. Similarly, the phase of nanomechanical parametric vibrations can be mapped to the states of a classical bit, enabling nanomechanical logic operations \cite{Mahboob2008} and the emulation of Ising spins \cite{Wang2013,McMahon2016,Heugel2022,Alvarez2024}. A prerequisite for the direct control of the phase states is the breaking of their amplitude symmetry, which can be realized by adding a nearly-resonant external drive to the parametric excitation \cite{Mahboob2010b,Ryvkine2006,Leuch2016}. Far away from resonance, the resonator responds mostly to this external drive, which means that the vibrational phases far below and far above resonance differ by $\pi$ radians. Adiabatically sweeping the angular frequency of the drive $\omega_\mathrm{d}$ and that of the pump $\omega_\mathrm{p}\simeq2\omega_\mathrm{d}$ through resonance ensures that the parametric resonator selects states with the phase imprinted by the external drive off resonance \cite{Mahboob2010b,Leuch2016,Nosan2019,Frimmer2019}. A striking phenomenon that accompanies an upward and downward frequency sweep is the appearance of a double hysteresis in the amplitude and in the phase of the vibrational response \cite{Leuch2016,Papariello2016,Supplemental_Material}. It originates from the nonlinear response to the compound excitation. It is within this double hysteresis that the two phase states can be individually addressed.

Low-dimensional nanomechanical resonators built on a chip, such as nanotubes, nanowires, nanobeams, nanoplates, and atomically-thin membranes \cite{Lemme2020,Xu2022}, possess two characteristics that are advantageous for parametric oscillators. Firstly, their resonant frequencies are highly tunable with a dc voltage applied to a nearby electrode \cite{Sazonova2004,Unterreithmeier2009}, which implies that $\omega_0$ can be efficiently modulated by electrical means to excite parametric modes \cite{Turner1998,Carr2000,Karabalin2010,Eichler2012,Mathew2016,Su2021}. Secondly, they feature pronounced conservative \cite{Sazonova2004,Aldridge2005,Kozinsky2006,Chen2009,Unterreithmeier2010} and dissipative mechanical nonlinearities \cite{Dykman1975,Lifshitz2009,Zaitsev2012,Eichler2011}, which favor the observation of the double hysteresis mentioned above \cite{Leuch2016,Papariello2016}. Among low-dimensional nanomechanical resonators, graphene resonators are generally interesting because their resonant frequencies are high, they dissipate little energy at low temperature, and they can be carved into large arrays \cite{vanderZande2010}. High resonant frequencies enable double hystereses whose jumps are separated clearly in frequency, making them easily identifiable. Low dissipation enables large-amplitude parametric oscillations beyond a moderate pump power. Large arrays may be developed into future networks of coupled parametric resonators to study Ising machines \cite{Wang2013,McMahon2016,Heugel2022,Alvarez2024}.

Here, we present a study of the parametric phase states of nanomechanical resonators based on few-layer graphene (FLG) at room temperature. We measure the amplitude and the phase of the vibrations of each resonator in response to a strong parametric pump combined with a weak external drive. We observe a double hysteresis in both responses as $\omega_0$ is detuned with a dc gate voltage. The double hysteresis allows us to control the phase of the parametric states. In the presence of an electrostatic force noise, the system is activated to jump between the states. We tune the stationary populations of these states by varying the phase difference between the external drive and the pump, the amplitude of the external drive, and the noise intensity. Our study is motivated by earlier works on symmetry breaking in parametric resonators based on nanobeams \cite{Mahboob2010b}, guitar strings \cite{Leuch2016}, electrical resonators \cite{Nosan2019}, levitated nanoparticles \cite{Frimmer2019}, and torsional resonators \cite{Han2024}. Our results contribute to this topic and demonstrate the breaking of the symmetry of the parametric phase states in graphene.

 \begin{figure}
 \includegraphics{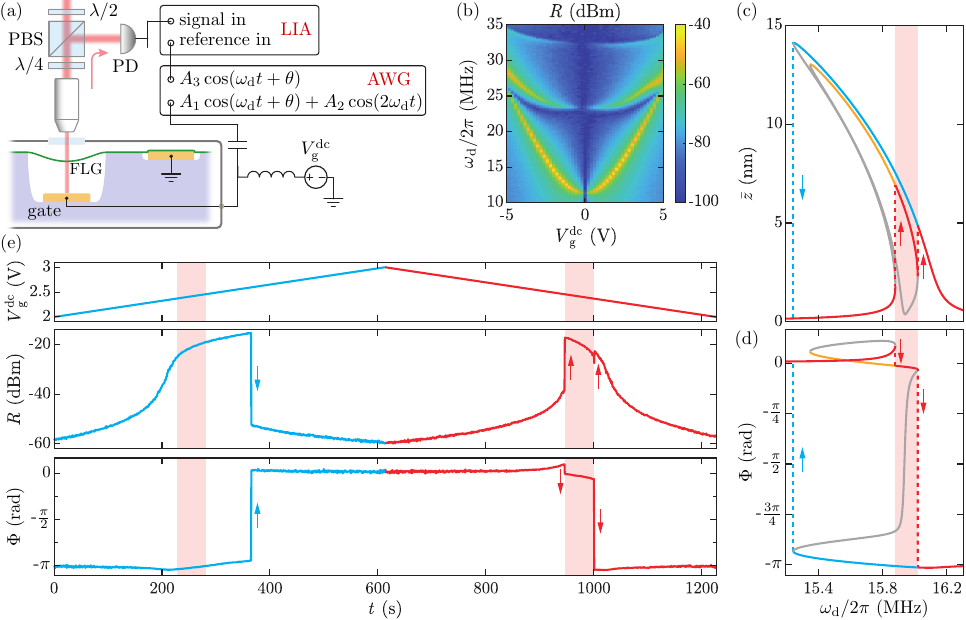}
 \caption{Large-amplitude parametric oscillations in a graphene nanomechanical resonator. (a) Measurement setup. PD: photodetector. PBS: polarizing beam splitter. $\lambda/2$, $\lambda/4$: half-wave, quarter-wave plate. FLG: few-layer graphene. LIA: lock-in amplifier. AWG: arbitrary waveform generator. The resonator is kept in vacuum and at room temperature. A laser beam with wavelength 633~nm is used. (b) Amplitude response of the resonator measured as an electrical power $R$ at the output of the photodetector as a function of drive frequency $\omega_\mathrm{d}$ and gate voltage $V_\mathrm{g}^\mathrm{dc}$. No parametric pump is applied. The power of the external drive applied to the gate is $-37$~dBm. (c) Amplitude response $\bar{z}$ and (d) phase response $\Phi$ of the stationary vibrations calculated as a function of $\omega_\mathrm{d}$ by solving Eq.~(\ref{EQM}) with $V_\mathrm{g}^\mathrm{dc}=2.4$~V, $A_1=4.2\times10^{-3}$~V, $A_2=0.635$~V, $\lambda=0.0465$, $\theta=125^\circ$, and the equation parameters listed in the text. Red, blue and orange traces depict stable states, while grey traces represent unstable states. The shaded region highlights the double hysteresis regime. Dotted lines and arrows indicate jumps upon changing $\omega_\mathrm{d}$. (e) $R$ and $\Phi$ measured at the output of the photodetector as $V_\mathrm{g}^\mathrm{dc}$ is increased (blue traces) and decreased (red traces). A trace with a certain color corresponds to the calculated branch with the same color in (c, d). Instead of changing $\omega_\mathrm{d}$, we change the resonant frequency $\omega_0$ with $V_\mathrm{g}^\mathrm{dc}$. The double hysteresis is highlighted by the shaded region. $A_1$, $A_2$, $\lambda$, and $\theta$ are the same as in (c, d).}     \label{setup}
 \end{figure}

Our FLG nanomechanical resonators are shaped as 3~$\mu$m-long ribbons (resonators~A, B, and C) and as a drum with a diameter of 3~$\mu$m (resonator D). They are suspended over a cavity etched in silicon dioxide, at the bottom of which a local gate electrode is patterned (Fig.~\ref{setup}a). The vertical distance between the flake and the top surface of the gate is 225~nm. Flexural vibrations of the flake are driven by applying a dc voltage $V_\mathrm{g}^\mathrm{dc}$ and an oscillating voltage $\delta V_\mathrm{g}(t)=A_1\cos(\omega_\mathrm{d}t+\theta)+A_2\cos(2\omega_\mathrm{d}t)$ between the gate and the flake, where $A_{1,2}$ are peak voltage amplitudes at the gate (accounting for the full reflection of the waveform at the gate). The term at $\omega_\mathrm{d}$ in $\delta V_\mathrm{g}$ is an external drive applied near the resonant frequency $\omega_0$ of the fundamental vibration mode. It creates a capacitive driving force of amplitude $F_\mathrm{d}\simeq C_\mathrm{g}^\prime V_\mathrm{g}^\mathrm{dc}A_1$, where $C_\mathrm{g}^\prime$ is the derivative of the capacitance $C_\mathrm{g}$ between the resonator and the gate with respect to the static displacement of the resonator. The term at $2\omega_\mathrm{d}$ is a parametric pump \cite{Eichler2012}. It leads to an effective modulation of the spring constant via three-wave mixing of the pump at $2\omega$ and the vibrations near $\omega_0$ in the presence of a nonlinear restoring force \cite{Eichler2023}. The phase difference between the external drive and the pump is $\theta$. Below, we present measurements performed on resonator~A. Data obtained with the other resonators are shown in Supplementary Information \cite{Supplemental_Material}. All measurements are done in a vacuum of $2\times10^{-6}$~mbar and at room temperature.

Our measurement setup is depicted in Fig.~\ref{setup}a. The resonator is placed in an optical standing wave formed between the gate electrode and a quarter-wave plate. The amount of optical energy the resonator absorbs is modulated as it vibrates, resulting in modulations in the intensity of the reflected light \cite{Mathew2016,Bunch2007,Barton2012,Davidovikj2016,Lu2021}. The latter are measured with a photodetector (PD), whose output $V_\mathrm{PD}(t)$ is a voltage that oscillates at the frequency of the mechanical vibrations and whose amplitude is proportional to the vibrational amplitude. $V_\mathrm{PD}(t)$ is measured with a radio-frequency lock-in amplifier (LIA) synchronized on a clock signal oscillating at $\omega_\mathrm{d}$ (Fig.~\ref{setup}a). Figure~\ref{setup}b displays the amplitude response of the resonator, $R=10\,\mathrm{log}_{10}[\langle(V_\mathrm{PD}(t)/2)^2\rangle/(50\times10^{-3})]$, as a function of $V_\mathrm{g}^\mathrm{dc}$ and $\omega_\mathrm{d}$ at $A_2=0$. $R$ is the power dissipated in the input impedance of the LIA in units of dBm; the factor of $1/2$ accounts for the voltage divider formed by the output impedance of the PD and the input impedance of the LIA, and $\langle\cdot\rangle$ averages over time. The data presented in Fig.~\ref{setup}b show that $\omega_0$, identified as the frequency of the peak of the response, is tunable with $V_\mathrm{g}^\mathrm{dc}$.

We model the vibrations using a nonlinear equation of motion for the center of mass of the resonator actuated by the external drive and the parametric pump \cite{Ryvkine2006,Mahboob2010b,Leuch2016,Papariello2016}. The equation reads:
\begin{equation}
\ddot{z}+\omega_0^2\left[1-\lambda\cos(\omega_\mathrm{p}t)\right]z+\tilde{\gamma}\dot{z}+\tilde{\alpha}z^3+\tilde{\eta}z^2\dot{z}=\tilde{F}_\mathrm{d}\cos(\omega_\mathrm{d}t+\theta)\,,\label{EQM}
\end{equation}
where $z$ is the amplitude of displacement in the flexural direction, $\tilde{\gamma}=\gamma/m_\mathrm{eff}$ with $\gamma$ the linear damping coefficient and $m_\mathrm{eff}$ the effective mass of the vibration mode, $\tilde{\alpha}=\alpha/m_\mathrm{eff}$ with $\alpha$ the parameter of the nonlinear conservative restoring force, $\tilde{\eta}=\eta/m_\mathrm{eff}$ with $\eta$ the parameter of the nonlinear dissipative force, and $\tilde{F}_\mathrm{d}=F_\mathrm{d}/m_\mathrm{eff}$. From the dimensions of resonator~A measured by atomic force microscopy, we estimate $m_\mathrm{eff}\simeq1.5\times10^{-16}$~kg. With the parametric pump turned off, we measure the frequency response of the resonator to the external drive. Setting $C_\mathrm{g}^\prime\simeq2.6\times10^{-9}$~F~m$^{-1}$ estimated with COMSOL and $\lambda=0$, we fit Eq.~(\ref{EQM}) to $v(\omega_\mathrm{d})=\kappa \bar{z}(\omega_\mathrm{d})$, where $v=[2\langle V_\mathrm{PD}^2(t)\rangle]^{1/2}$ is the peak amplitude of $V_\mathrm{PD}(t)$, $\bar{z}$ is the steady-state vibrational amplitude, and $\kappa$ is a linear transduction factor from meter to Volt. From this fit, we extract $\tilde{\gamma}\simeq2.05\times10^6$~rad/s, $\alpha\simeq-9.5\times10^{14}$~kg~m$^{-2}$~s$^{-2}$, and $\eta\simeq1.3\times10^6$~kg~m$^{-2}$~s$^{-1}$ (see Supplementary Information~\cite{Supplemental_Material}). The modulation depth $\lambda$ can be quantified from the response of the resonator to the pump in the absence of an external drive, $A_1=0$, as \cite{Lifshitz2009,Eichler2012,Supplemental_Material}
\begin{equation}
\lambda=\frac{2A_2}{A_{2,\mathrm{th}}}\frac{\tilde{\gamma}}{\omega_0}\simeq A_2\times7.4\times10^{-2}\,\,\mathrm{V}^{-1}\,,\label{lambda}
\end{equation}
where $A_{2,\mathrm{th}}$ is the threshold value of the pump amplitude beyond which a response is measured, and $\omega_0/2\pi\simeq16\times10^6$~Hz at $V_\mathrm{g}^\mathrm{dc}=2.4$~V.

Armed with the above parameters, we solve Eq.~(\ref{EQM}) using the perturbative averaging method \cite{Leuch2016}. The calculated amplitude $\bar{z}$ and the calculated phase $\Phi$ (with respect to the external drive) of the stationary vibrations are shown as a function of $\omega_\mathrm{d}$ in Figs.~\ref{setup}c, d for $A_1=4.2\times10^{-3}$~V, $A_2=0.635$~V, $V_\mathrm{g}^\mathrm{dc}=2.4$~V, $\omega_0/2\pi=16\times10^6$~Hz, and $\theta=125^\circ$. At first sight, both responses resemble those of a nonlinear resonator in the absence of a parametric pump, with an upper amplitude branch and a lower amplitude branch that form a shark fin, as well as $\Phi\simeq0$ below and $\Phi\simeq-\pi$ above resonance. However, the interplay of nonlinearities, external drive, and pump gives rise to additional states near resonance, which become apparent as $\omega_\mathrm{d}$ is varied. Decreasing $\omega_\mathrm{d}/2\pi$ from 16.2~MHz, the resonator enters the parametric regime and settles in the largest amplitude state (blue trace in Fig.~\ref{setup}c) with $\Phi\simeq-\pi$, until it jumps to a low amplitude state with $\Phi\simeq0$ (vertical blue arrows in Figs.~\ref{setup}c, d). Upon reversing the sweep direction and increasing $\omega_\mathrm{d}$, the resonator jumps to an intermediate amplitude state with $\Phi\simeq0$ (leftmost vertical red arrow in Figs.~\ref{setup}c, d). Past this point, decreasing $\omega_\mathrm{d}$ again would take the resonator along the orange branches in Figs.~\ref{setup}c, d, where $\bar{z}$ increases and $\Phi\simeq0$. Alternatively, continuing to increase $\omega_\mathrm{d}$ makes the resonator jump to the largest amplitude state with $\Phi\simeq-\pi$ (rightmost vertical red arrow in Figs.~\ref{setup}c, d). If $\omega_\mathrm{d}$ is first decreased from far above resonance and then increased from far below resonance, the frequency range bound by these two jumps defines a double hysteresis that makes it possible to choose the phase of vibrations. The two phase states, represented by the two traces in red and blue within the shaded region in Figs.~\ref{setup}c, d, can be addressed individually, now that their amplitude symmetry has been broken by the weak external drive.  

Guided by our calculations, we experimentally prepare the resonator in one of the two phase states. To quantify the response, we measure the power $R$ and the phase $\Phi$ of $V_\mathrm{PD}(t)$. $R$ is a measure of $\bar{z}^2$ since $V_\mathrm{PD}(t)\propto z(t)$. Attempts at measuring $R$ and $\Phi$ as a function of $\omega_\mathrm{d}$ have proven challenging; we have found that sweeping $\omega_\mathrm{d}$ creates stochastic jumps in $R$ and $\Phi$ (see Supplementary Information~\cite{Supplemental_Material}). Instead of sweeping $\omega_\mathrm{d}$, we detune the resonant frequency with $V_\mathrm{g}^\mathrm{dc}$ to bring the resonator in and out of parametric resonance \cite{Nosan2019}, having set $A_1$, $A_2$, $\lambda$, and $\theta$ as in Figs.~\ref{setup}c, d. Figure~\ref{setup}e shows $R$ and $\Phi$ as $V_\mathrm{g}^\mathrm{dc}$ is first increased and then decreased. Upon decreasing $V_\mathrm{g}^\mathrm{dc}$, we observe two jumps in $R$ and two concomitant jumps in $\Phi$. Interestingly, while the second jump in $R$ is barely noticeable (rightmost arrow in Fig.~\ref{setup}e), the corresponding change in $\Phi$ is large. This behavior is reproduced by our calculations, where the two vibrational states have a similarly large amplitude but a phase difference of nearly $\pi$ radians \cite{Mahboob2010b,Leuch2016,Papariello2016}. Within the double hysteresis (highlighted by the shaded regions in Fig.~\ref{setup}e), the phase of vibrations is controllably $\simeq0$ (red trace in Fig.~\ref{setup}e) or $\simeq-\pi$ (blue trace in Fig.~\ref{setup}e).

\begin{figure}[t]
 \includegraphics{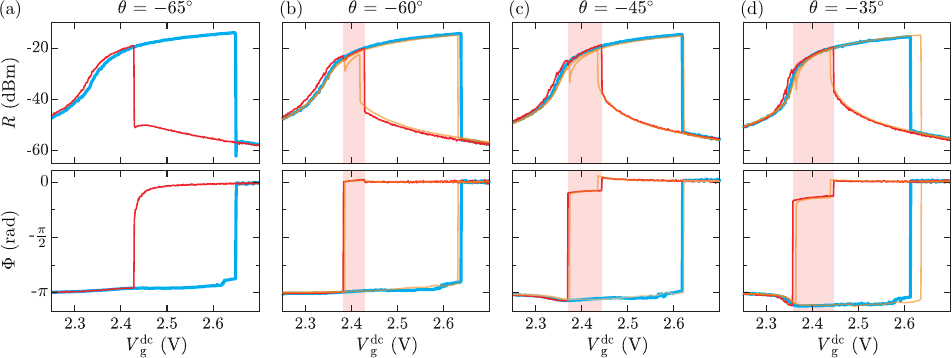}
 \caption{Measured amplitude and phase of the output signal of the photodetector as a function of $V_\mathrm{g}^\mathrm{dc}$ for various values of $\theta$. $A_1=4.2\times10^{-3}$~V, $A_2=0.635$~V, and $\lambda=0.0465$, as in Figs.~\ref{setup}c-e. Blue (red) traces: increasing (decreasing) $V_\mathrm{g}^\mathrm{dc}$. From (a) to (d): $\theta=-65^\circ$, $-60^\circ$, $-45^\circ$, $-35^\circ$. Traces in orange are fits of Eq.~(\ref{EQM}) where the only fit parameter is the transduction factor $\kappa$.}     \label{theta}
\end{figure}

As expected from Eq.~(\ref{EQM}), we find that the existence of the double hysteresis strongly depends on $\theta$, the phase difference between the parametric pump and the external drive. Figure~\ref{theta} shows $R$ and $\Phi$ measured as a function of $V_\mathrm{g}^\mathrm{dc}$ for several values of $\theta$. $A_1$, $A_2$ and $\lambda$ are the same as in Figs.~\ref{setup}c-e. Blue (red) traces are obtained upon increasing (decreasing) $V_\mathrm{g}^\mathrm{dc}$. While the double hysteresis is not observed for $\theta=-65^\circ$ (Fig.~\ref{theta}a), it is visible for $\theta=-60^\circ$ and its width increases as $\theta$ is changed from $-60^\circ$ to $-35^\circ$ (Figs.~\ref{theta}b-d). However, it disappears again for $\theta=-30^\circ$ (see Supplementary Information~\cite{Supplemental_Material}). As in Fig.~\ref{setup}e, the jump measured in $R$ between the two phase states is barely detectable, while the concomitant jump in $\Phi$ is large. We can reproduce our data with the model described by Eq.~(\ref{EQM}), where the only fit parameter is the transduction factor $\kappa$. Our model does not include the mode near $2\omega_0$ that we observe in Fig.~\ref{setup}b. To confirm that this mode does not play a significant role, we have conducted similar studies with resonators~C and D, which do not feature such a mode. We have found that their measured responses show the same qualitative features as in resonator~A and can be reproduced by our model~\cite{Supplemental_Material}.

\begin{figure}
\includegraphics{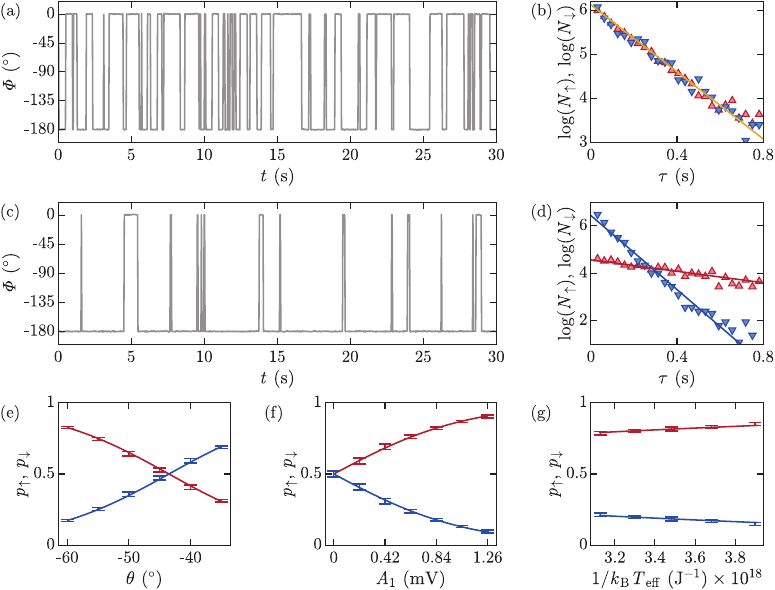}
\caption{Stochastic transitions and occupation probabilities of symmetry-broken parametric phase states in the presence of a white Gaussian force noise. (a) Stochastic jumps in the phase $\Phi$ of the output signal of the photodetector measured as a function of time and referenced at the input of the lock-in amplifier, with $A_1=0$ and $k_\mathrm{B}T_\mathrm{eff}\simeq2.2\times10^{-19}$~J. (b) Distribution of the residence times $\tau$ in states $\uparrow$ and $\downarrow$ extracted from $\Phi(t)$ data. $N_{\downarrow}$ is the number of transitions out of state $\downarrow$ (number of jumps from $\Phi\simeq-\pi$ to $\Phi\simeq0$, red markers) and $N_{\uparrow}$ is the number of transitions out of state $\uparrow$ (number of jumps from $\Phi\simeq0$ to $\Phi\simeq-\pi$, blue markers), after the resonator has resided in a state for a time $\tau$. The yellow trace is a fit of the Poisson distribution. (c) Stochastic jumps in $\Phi$ measured with $A_1= 1.05\times10^{-3}$~V, $\theta=-60^\circ$, and $k_\mathrm{B}T_\mathrm{eff}\simeq2.2\times10^{-19}$~J. (d) Distribution of $\tau$ in states $\uparrow$ and $\downarrow$ extracted from $\Phi(t)$ data. Blue and red traces are fits of the Poisson distribution to the data. (e-g) Stationary occupation probabilities $p_\downarrow$ and $p_\uparrow$ of states $\downarrow$ and $\uparrow$ as a function of $\theta$ (with $A_1= 1.05\times10^{-3}$~V and $k_\mathrm{B}T_\mathrm{eff}\simeq2.6\times10^{-19}$~J), $A_1$ (with $\theta=-60^\circ$ and $k_\mathrm{B}T_\mathrm{eff}\simeq2.2\times10^{-19}$~J), and $1/k_\mathrm{B}T_\mathrm{eff}$ (with $A_1= 1.05\times10^{-3}$~V and $\theta=-60^\circ$). Solid traces are fits of Eq.~(\ref{psteady}). Fit parameters are $\chi\simeq1.1\times10^{-7}$~J~N$^{-1}$ and $\delta\simeq-46.5^\circ$ in (e), $\chi\cos(\theta+\delta)\simeq-3.1\times10^{-8}$~J~N$^{-1}$ in (f), and $\chi\cos(\theta+\delta)\simeq-3.2\times10^{-8}$~J~N$^{-1}$ in (g). $V_\mathrm{g}^\mathrm{dc}=2.4$~V, $\omega_0/2\pi\simeq\omega_\mathrm{d}/2\pi=16\times10^6$~Hz, and $A_2=0.635$~V in all panels.}\label{occupation}     
\end{figure}

The symmetry breaking of the phase states is also apparent in their occupation probabilities in the presence of a weak broadband force noise. This force noise is created by applying a white Gaussian voltage noise to the gate; it is quantified as an effective vibrational temperature $T_\mathrm{eff}$ using the thermal vibrations measured at room temperature as a calibration \cite{Supplemental_Material}. Figure~\ref{occupation} displays $\Phi$ measured as a function of time in the presence of the force noise, without applying an external drive ($A_1=0$). We have set $A_2=0.635$~V, $V_\mathrm{g}^\mathrm{dc}=2.4$~V, and $\omega_\mathrm{d}/2\pi=16\times10^6$~Hz. $\Phi$ jumps stochastically between $\simeq0$ and $\simeq-\pi$, indicating random and uncorrelated transitions between the two states \cite{Dykman1998,Lapidus1999,Marthaler2006,Kim2006,Chan2007,Mahboob2010b,Han2024}. The stationary occupation probability of each state is measured as $p_{\uparrow,\downarrow}=t_{\uparrow,\downarrow}/t_\mathrm{meas}$, where $\uparrow$ ($\downarrow$) labels the state with $\Phi\simeq0$ ($\Phi\simeq-\pi$), $t_\uparrow$ ($t_\downarrow$) is the total time spent in state $\uparrow$ ($\downarrow$), and $t_\mathrm{meas}$ is the total measurement time. We find $p_\uparrow\simeq p_\downarrow$, showing that the states are degenerate, in agreement with Ref.~\cite{Mahboob2010b}. The statistics of transitions is well described by the Poisson distribution \cite{Dolleman2019}, as shown in Fig.~\ref{occupation}b where the number of transitions out of a given state $N_{\uparrow,\downarrow}$ is plotted as a function of the residence time $\tau$. A fit of an exponential decay allows us to estimate the transition rates $W_\uparrow$ out of state $\uparrow$ and $W_\downarrow$ out of state $\downarrow$, where we find $W_\uparrow\simeq W_\downarrow$. Measuring $W_{\uparrow,\downarrow}$ as a function of $T_\mathrm{eff}$, we verify that they follow an activation law \cite{Dykman1998,Lapidus1999,Marthaler2006,Kim2006,Chan2007,Han2024}:
\begin{equation}
W_{\uparrow,\downarrow}=C\exp\left(-U_{\uparrow,\downarrow}/k_\mathrm{B}T_\mathrm{eff}\right)\,,\label{W}
\end{equation}
with $C$ a constant and $U_\uparrow$ ($U_\downarrow$) the activation energy out of state $\uparrow$ ($\downarrow$), where $U_\uparrow\simeq U_\downarrow\simeq3.6\times10^{-18}$~J (see Supplementary Information~\cite{Supplemental_Material}). It is interesting to compare $T_\mathrm{eff}$ in our work with that estimated in Ref.~\cite{Dolleman2019}, where stochastic switching between the two stable states of a graphene resonator driven into the nonlinear regime with a strong external drive (without a parametric pump) is measured. Both resonator~A and the resonator in Ref.~\cite{Dolleman2019} have a similar $\omega_0$ and a similar interstate activation energy. In our work, at the largest value of $T_\mathrm{eff}\simeq2.3\times10^4$~K we estimate $W_{\uparrow,\downarrow}\simeq35$~Hz. The authors of Ref.~\cite{Dolleman2019} explore higher transition rates $r_k$, hence their values of $T_\mathrm{eff}$ are generally higher than ours. Yet, they estimate $r_k\simeq35$~Hz near $T_\mathrm{eff}\simeq3.5\times10^4$~K, which is a somewhat higher temperature than our estimate.

Applying a small external drive lifts the occupation degeneracy of the states. We prepare the resonator within the frequency range between the jumps using $A_1=1.05\times10^{-3}$~V, $A_2=0.635$~V,  $\theta=-60^\circ$, and $V_\mathrm{g}^\mathrm{dc}=2.4$~V. $\Phi$ displays jumps that favor state $\downarrow$ (that is, $\Phi\simeq-\pi$ is observed most often, see Fig.~\ref{occupation}c). Accordingly, the statistics of the transitions reveals that $W_\downarrow\ll W_\uparrow$ (Fig.~\ref{occupation}d). We then measure the occupation probabilities of the two states as a function of $\theta$ and $A_1$, the parameters of the symmetry-breaking drive. With $A_1=1.05\times10^{-3}$~V, $p_\uparrow$ and $p_\downarrow$ show a clear dependence on $\theta$ (Fig.~\ref{occupation}e). Similarly, with $\theta=-60^\circ$, $p_\uparrow$ and $p_\downarrow$ strongly vary with $A_1$ (Fig.~\ref{occupation}f). Both behaviors can be explained by the change in transition rates caused by the external drive $F_\mathrm{d}\cos(\omega_\mathrm{d}t+\theta)$. According to the theory in Ref.~\cite{Ryvkine2006}, the external drive modifies the activation energies as
\begin{equation}
U_{\uparrow,\downarrow}=\bar{U}+\sigma_{\uparrow,\downarrow}\chi F_\mathrm{d}\cos(\theta+\delta)\,,\label{logsusceptibility}
\end{equation}
where $\bar{U}$ is the activation energy in the absence of a symmetry breaking drive, $\sigma_\uparrow=1$, $\sigma_\downarrow=-1$, and $\chi$ and $\delta$ are the magnitude and the phase of the logarithmic susceptibility of the resonator. Expressing $p_{\uparrow,\downarrow}$ as the stationary solutions to the balance equation for the occupation probabilities,
\begin{equation}
p_{\uparrow,\downarrow}=\frac{W_{\downarrow,\uparrow}}{W_\uparrow+W_\downarrow}\,,
\end{equation}
Eqs.~(\ref{W}) and (\ref{logsusceptibility}) yield \cite{Han2024}
\begin{equation}
p_{\uparrow,\downarrow}=\left\{1+\exp\left[\frac{-2\sigma_{\uparrow,\downarrow}\chi F_\mathrm{d}\cos(\theta+\delta)}{k_\mathrm{B}T_\mathrm{eff}}\right]\right\}^{-1}\,.\label{psteady}
\end{equation}
We fit Eq.~(\ref{psteady}) to our measurements in Figs.~\ref{occupation}e, f (solid traces), using $\chi$ and $\delta$ as fit parameters. With $A_1=1.05\times10^{-3}$~V and $\theta=-60^\circ$, we also measure $p_{\uparrow,\downarrow}$ as a function of $1/k_\mathrm{B}T_\mathrm{eff}$, and fit Eq.~(\ref{psteady}) to the data in Fig.~\ref{occupation}g. Measurements of the logarithmic susceptibility of resonator~A, along with related measurements performed on resonators~C and D, are shown in Supplementary Information~\cite{Supplemental_Material}.

We have demonstrated that parametric phase states with broken amplitude symmetry can be created in nanomechanical resonators made from graphene. This approach differs from earlier methods where the vibrational phase is manipulated by coherently cycling the populations of the normal modes of strongly coupled resonators with a radio-frequency field \cite{Okamoto2013,Faust2013,ZZZhang2020}. Looking ahead, we envision the realization of phase logic devices with a combination of useful properties. (i) Our resonator has a footprint of a few squared micrometers, which allows massive parallel implementation on a single chip. (ii) Graphene resonators have a low mass, and can therefore be driven with a low power. (iii) Our device has a bandwidth of about 1 MHz, which should make it possible to flip the phase states within a few microseconds using optimized protocols. (iv) Graphene resonators can be tuned over a wide frequency range with electrical gates. Such tuning is not only important to obtain resonant coupling between neighboring devices, but, as we show here, provides a toolbox for flipping the phase states of individual devices in a deterministic manner. The above properties hold promise for large on-chip networks of highly tunable parametric oscillators. Such networks could become a resource for solving complex optimization tasks. The combination of small size, low power, high bandwidth, and frequency tuning and phase flipping via gate electrodes makes graphene resonators a promising platform for this endeavor, provided that issues such as nonuniform strain distribution and defects within the membrane can be mitigated.

\begin{acknowledgments}
This work was supported by the National Natural Science Foundation of China (grant numbers 62150710547 and 62074107) and the project of the Priority Academic Program Development (PAPD) of Jiangsu Higher Education Institutions. The authors are grateful to Prof. Wang Chinhua for his strong support. J.M. is grateful to Antoine Reserbat-Plantey and Fabien Vialla for inspiring discussions.
\end{acknowledgments}

\section*{\textbf{Supplementary Information}}

\section{Response to the external drive, to the parametric pump, and to the compound excitation}

\begin{figure}[h]
\includegraphics[width=12.5cm]{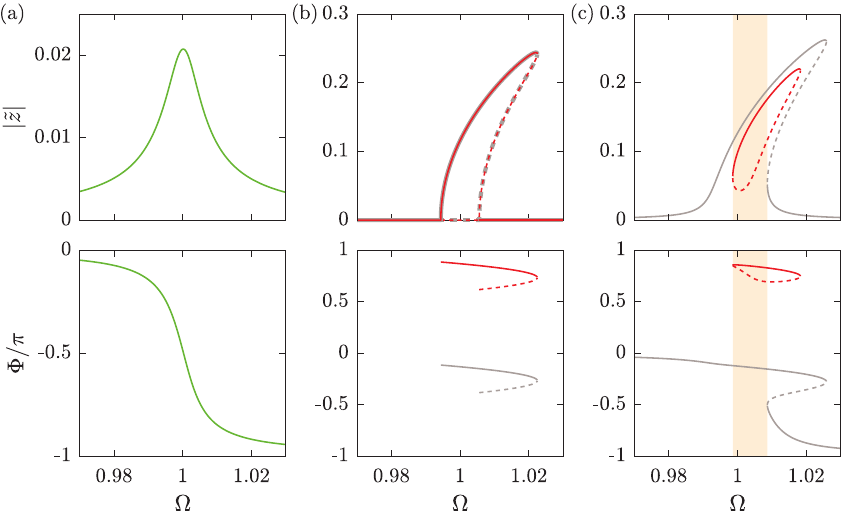}
\caption{Response to the external drive, to the parametric pump, and to the compound excitation. (a) Peak amplitude of external drive $F_\mathrm{d}=3\times10^{-11}$~N. No parametric pump is applied. (b) Parametric pump with spring constant modulation depth $\lambda=0.03$. No external drive is applied. (c) Compound excitation with $F_\mathrm{d}=3\times10^{-11}$ and $\lambda=0.03$. The phase difference between the external drive and the parametric pump is $\theta=\pi/4$. The shaded area highlights the region of the double hysteresis (see main text). Solid (dashed) traces represent stable (unstable) solutions. In all panels, $m_\mathrm{eff}=6\times10^{-17}$~kg, $\omega_0/2\pi=50\times10^6$~Hz, the quality factor $Q=100$, $\alpha=10^{16}$~kg~m$^{-2}$~s$^{-2}$, and the parameter of the nonlinear dissipative force $\eta=10^7$~kg~m$^{-2}$~s$^{-1}$.} \label{SFig0}
\end{figure}

Figure~\ref{SFig0} shows the amplitude $\vert\tilde{z}\vert$ and the phase $\Phi$ of the solution to Eq.~(1) in the main text. Figure~\ref{SFig0}a shows the response to a weak external drive in the absence of a parametric pump. The response is linear. Figure~\ref{SFig0}b shows the response to a large parametric pump above threshold in the absence of an external drive. The response is nonlinear. Two large-amplitude stable solutions with the same amplitude and a phase difference of $\pi$~radians are observed. Figure~\ref{SFig0}c shows the response to a compound excitation composed of a weak external drive and a large parametric pump above threshold. The response is nonlinear. The amplitude symmetry of the two large-amplitude stable states is broken. Within the shaded area that highlights the double hysteresis region, the phase difference between these states is nearly $\pi$~radians.

Following Ref.~\cite{Papariello2016}, the vibrational amplitude is here normalized as $\vert\tilde{z}\vert=\bar{z}\sqrt{\alpha/(m_\mathrm{eff}\omega_0^2)}$, where $\bar{z}$ is the steady-state vibrational amplitude (in units of m), $\alpha$ is the parameter of the nonlinear conservative restoring force (in units of kg~m$^{-2}$~s$^{-2}$), $m_\mathrm{eff}$ is the effective mass of the mode (in units of kg) and $\omega_0$ is the angular resonant frequency of the mode (in units of rad~s$^{-1}$). The response is plotted as a function of the normalized frequency $\Omega=\omega_\mathrm{d}/\omega_0$, where $\omega_\mathrm{d}$ is the angular frequency of the external drive (Figs.~\ref{SFig0}a, c) or as a function of $\Omega=\omega_\mathrm{p}/(2\omega_0)$, where $\omega_\mathrm{p}$ is the angular frequency of the parametric pump (Fig.~\ref{SFig0}b). The parameters are listed in the captions to Fig.~\ref{SFig0} and are similar to those employed in the study of resonator~D with a sign change for $\alpha$ (Supplementary Note~\ref{responseofresonatorD}).

\section{Estimating the parameters of the equation of motion of resonator~A}\label{EQMsection}

In the absence of a parametric pump, $\lambda=0$, the equation of motion for the steady-state displacement of the center of mass of the resonator reads:

\begin{equation}
\rho^3\left[\frac{9}{16}\tilde{\alpha}^2+\frac{1}{16}\tilde{\eta}^2\omega_\mathrm{d}^2\right]+\rho^2\left[-\frac{3}{2}\tilde{\alpha}\left(\omega_\mathrm{d}^2-\omega_0^2\right)+\frac{1}{2}\tilde{\eta}\tilde{\gamma}\omega_\mathrm{d}^2\right]+\rho\left[\left(\omega_\mathrm{d}^2-\omega_0^2\right)^2+\tilde{\gamma}^2\omega_\mathrm{d}^2\right]-\tilde{F}_\mathrm{d}^2=0\,,\label{EQM}
\end{equation} 
where $\rho=\bar{z}^2$, with $\bar{z}$ the steady-state amplitude of displacement, and where $\tilde{\alpha}=\alpha/m_\mathrm{eff}$, $\tilde{\eta}=\eta/m_\mathrm{eff}$, $\tilde{\gamma}=\gamma/m_\mathrm{eff}$, $\tilde{F}_\mathrm{d}=F_\mathrm{d}/m_\mathrm{eff}$, and $\omega_\mathrm{d}=2\pi f_\mathrm{d}$. $\alpha$ is the parameter of the nonlinear conservative restoring force, $\eta$ is the parameter of the nonlinear dissipative force, $\gamma$ is the linear damping coefficient, $m_\mathrm{eff}$ and $\omega_0$ are the effective masse and the angular resonant frequency of the vibration mode, respectively, and $F_\mathrm{d}$ and $\omega_\mathrm{d}$ are the amplitude and the angular frequency of the external drive, respectively.

We consider resonator~A in the main text and measure the power $R$ of the signal at the output of the photodetector as a function of the drive frequency $f_\mathrm{d}=\omega_\mathrm{d}/2\pi$. Labeling the output voltage of the photodetector as $V_\mathrm{PD}(t)$, the power in units of dBm that is dissipated across the input impedance of the lock-in amplifier reads $R=10\,\mathrm{log}_{10}[\langle(V_\mathrm{PD}(t)/2)^2\rangle/(50\times10^{-3})]$, where the factor of 1/2 accounts for the voltage divider formed by the output impedance of the photodetector and the input impedance of the lock-in amplifier, and $\langle\cdot\rangle$ averages over time $t$. Our vibration detection scheme is linear, $V_\mathrm{PD}(t)=\kappa z(t)$, with $\kappa=P_\mathrm{inc}\times\mathrm{T}\times\mathrm{G}\times\vert\mathrm{d}r/\mathrm{d}z_\mathrm{s}\vert$ a transduction factor from meter to Volt; $P_\mathrm{inc}$ is the optical power incident on the resonator, $\mathrm{T}$ is the transmittance of the optical path from the resonator to the photodetector, $\mathrm{G}$ is the transimpedance gain of the photodetector, and $\mathrm{d}r/\mathrm{d}z_\mathrm{s}$ is the derivative with respect to the static displacement of the resonator $z_\mathrm{s}$ of the optical reflection coefficient $r$ at the surface of the resonator facing the light source. The peak amplitude of the external drive reads $F_\mathrm{d}\simeq C_\mathrm{g}^\prime V_\mathrm{g}^\mathrm{dc}\delta V_\mathrm{g}$, where $V_\mathrm{g}^\mathrm{dc}$ and $\delta V_\mathrm{g}$ are the dc voltage and the peak amplitude of the oscillating voltage applied between the gate and the resonator, respectively, and $C_\mathrm{g}^\prime=\mathrm{d}C_\mathrm{g}/\mathrm{d}z_\mathrm{s}$ is the derivative of the capacitance between the gate and the resonator. $\delta V_\mathrm{g}$ accounts for the full reflection of the voltage wave at the gate, hence $\delta V_\mathrm{g}$ is twice the amplitude of the voltage wave incident on the gate. The power of the voltage wave driving the resonator that would be dissipated across a 50~Ohm resistor is $P_\mathrm{d}=10\,\mathrm{log}_{10}[\delta V_\mathrm{g}^2/(2\times50\times10^{-3})]$ in units of dBm, where the factor of 1/2 accounts for the rms averaging of the voltage. We estimate $C_\mathrm{g}^\prime\simeq2.6\times10^{-9}$~F~m$^{-1}$ using COMSOL. Finally, we estimate $m_\mathrm{eff}\simeq1.5\times10^{-16}$~kg from atomic force microscopy images of the few-layer graphene membrane.

\begin{figure}[h]
\includegraphics{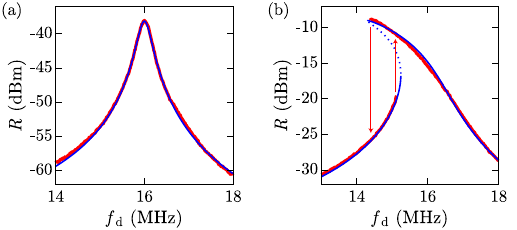}
\caption{Estimating the parameters of the equation of motion for resonator~A. Red traces are measured data and blue traces are fits of Eq.~(\ref{EQM}) [blue dots indicate unstable solutions]. (a) Linear response. $F_\mathrm{d}\simeq2.8\times10^{-11}$~N; $P_\mathrm{inc}\simeq15.5\times10^{-6}$~W. (b) Nonlinear response. $F_\mathrm{d}\simeq8.9\times10^{-10}$~N; $P_\mathrm{inc}\simeq19\times10^{-6}$~W. In (a) and (b), $V_\mathrm{g}^\mathrm{dc}=2.4$~V, $\mathrm{G}=2.5\times10^5$~V~W$^{-1}$; $\mathrm{T}\simeq0.5$; $\vert\mathrm{d}r/\mathrm{d}z_\mathrm{s}\vert\simeq4.5\times10^{6}$~m$^{-1}$.} \label{paraEQM}
\end{figure}

Figure~\ref{paraEQM}a shows $R(f_\mathrm{d})$ in response to a weak drive, for which the response is linear. Fitting Eq.~(\ref{EQM}) to our measured data, we estimate $\tilde{\gamma}\simeq2.05\times10^6$~rad~s$^{-1}$. Figure~\ref{paraEQM}b shows $R(f_\mathrm{d})$ in response to a larger drive, for which the response is nonlinear. Fitting Eq.~(\ref{EQM}) to the data, we estimate $\alpha\simeq-9.5\times10^{14}$~kg~m$^{-2}$~s$^{-2}$ and $\eta\simeq1.3\times10^6$~kg~m$^{-2}$~s$^{-1}$.

\section{Estimating the spring constant modulation depth $\lambda$ of resonator~A}

Applying a parametric pump voltage $A_2\cos(2\omega_\mathrm{d}t)$ with $\omega_\mathrm{d}\simeq\omega_0$, where $\omega_0$ is the angular resonant frequency, effectively modulates the spring constant of the resonator. The modulation depth $\lambda$ can be estimated from the response of the resonator to the pump in the absence of an external drive, $F_\mathrm{d}=0$, as \cite{Lifshitz2009,Eichler2012}
\begin{equation}
\lambda=\frac{2A_2}{A_{2,\mathrm{th}}}\frac{\tilde{\gamma}}{\omega_0}\,,\label{lambda}
\end{equation}
where $A_{2,\mathrm{th}}$ is the threshold value of the pump amplitude beyond which a response is measured.

We consider resonator~A in the main text. Figure~\ref{lambdafig} shows the power $R$ of the signal at the output of the photodetector (see Supplementary Note~1 for the definition of $R$) as a function of $A_2$ upon increasing (Fig.~\ref{lambdafig}a) and decreasing (Fig.~\ref{lambdafig}b) $f_\mathrm{d}=\omega_\mathrm{d}/2\pi$. We have set $V_\mathrm{g}^\mathrm{dc}=2.4$~V, yielding $\omega_0/2\pi\simeq16\times10^6$~Hz. We find that $R$ becomes measurable above $A_{2,\mathrm{th}}\simeq0.555$~V. Hence we estimate $\lambda\simeq A_2\times7.4\times10^{-2}$~V$^{-1}$.

\begin{figure}[h]
\includegraphics{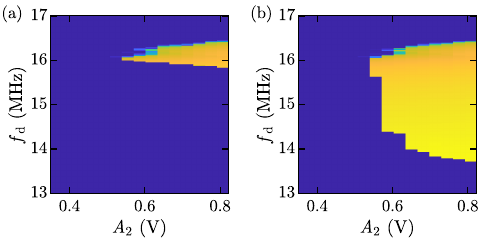}
\caption{Estimating the spring constant modulation depth $\lambda$ of resonator~A. A pump voltage $A_2\cos(2\times2\pi f_\mathrm{d}t)$ is applied between the resonator and the gate. The power of the signal at the output of the photodetector $R$ is shown as a function of $A_2$ upon increasing (a) and decreasing (b) $f_\mathrm{d}$. Gate voltage $V_\mathrm{g}^\mathrm{dc}=2.4$~V. Power scale: blue, $-80$~dBm; yellow, $-6$~dBm.} \label{lambdafig}
\end{figure}

\section{Stochastic frequency jumps upon changing $f_\mathrm{d}$ in resonator~B}

\begin{figure}[h]
\includegraphics{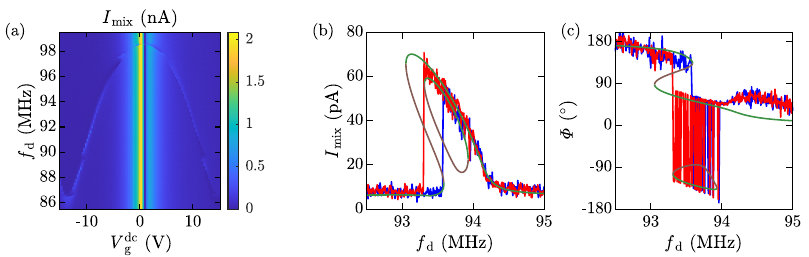}
\caption{Stochastic frequency jumps in the response of resonator~B upon changing $f_\mathrm{d}$ in the presence of a parametric pump at 3~K. A two-source mixing technique is used to measure vibrations. (a) Amplitude of the mixing current $I_\mathrm{mix}$ at the drain electrode of the resonator as a function of gate voltage $V_\mathrm{g}^\mathrm{dc}$ and drive frequency $f_\mathrm{d}$ in the absence of a parametric pump. $A_1=8.5\times10^{-3}$~V. (b) Amplitude $I_\mathrm{mix}$ and (c) phase of the mixing current $\Phi$ upon incrementing (blue traces) and decrementing (red traces) $f_\mathrm{d}$. $V_\mathrm{g}^\mathrm{dc}=-8$~V; $A_1=7\times10^{-3}$~V; $A_2 =0.05$~V; $\theta=0^\circ$. The green traces are stable solutions of Eq.~(1) in the main text, while the brown traces are unstable solutions. The data analysis is based on $m_\mathrm{eff}\simeq3\times10^{-17}$~kg, $\tilde{\gamma}\simeq2.96\times10^5$~rad~s$^{-1}$, $\alpha\simeq-2.4\times10^{17}$~kg~m$^{-2}$~s$^{-2}$, and $\eta\simeq3.2\times10^8$~kg~m$^{-2}$~s$^{-1}$.} \label{cryo}
\end{figure}

We measured resonator~B at a cryostat temperature of 3~K. A two-source mixing technique was used to measure vibrations, instead of the optical technique employed in the main text and elsewhere in these Supplementary Notes. Figure~\ref{cryo}a depicts the amplitude of the electromechanical mixing current $I_\mathrm{mix}$ in the linear regime of vibrations as a function of the gate voltage $V_\mathrm{g}^\mathrm{dc}$ and of the drive frequency $f_\mathrm{d}$, in the absence of a parametric pump, showing that the resonant frequency is still tunable at low temperature. With the compound excitation $\delta V_\mathrm{g}(t)=A_1\cos(2\pi f_\mathrm{d}t+\theta)+A_2\cos(2\times2\pi f_\mathrm{d}t)$ applied between the gate and the resonator, attempts at measuring $I_\mathrm{mix}$ and the phase of the mixing current $\Phi$ as a function of $f_\mathrm{d}$ proved challenging. The waveform $\delta V_\mathrm{g}(t)$ could not be changed adiabatically as $f_\mathrm{d}$ was stepped, creating stochastic jumps in $I_\mathrm{mix}$ and in $\Phi$.  Figures~\ref{cryo}b and \ref{cryo}c show $I_\mathrm{mix}$ and $\Phi$, respectively, as $f_\mathrm{d}$ is being incremented and then decremented by small steps. Stochastic jumps are observed in both responses. However, it is still possible to fit Eq.~(1) in the main text to the data, revealing the double hysteresis that signals the breaking of the symmetry of the parametric phase states.

\section{Amplitude and phase responses of resonator~A as a function of $V_\mathrm{g}^\mathrm{dc}$ at $\theta=-30^\circ$}

\begin{figure}[h]
\includegraphics{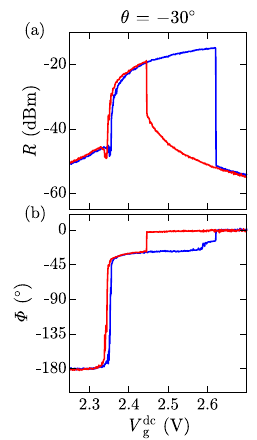}
\caption{Amplitude and phase responses, $R$ and $\Phi$, of resonator~A as a function of gate voltage $V_\mathrm{g}^\mathrm{dc}$ at $\theta=-30^\circ$. $A_1=4.2\times10^{-3}$~V, $A_2=0.635$~V, and $\lambda=0.0465$.} \label{thetam30deg}
\end{figure}

Figure~\ref{thetam30deg} complements Fig.~2 in the main text. It shows the amplitude and the phase responses, $R$ and $\Phi$, of resonator~A as a function of gate voltage $V_\mathrm{g}^\mathrm{dc}$ at $\theta=-30^\circ$, where $\theta$ is the phase difference between the external drive and the parametric pump. The following parameters are used: amplitude of external drive $A_1=4.2\times10^{-3}$~V, amplitude of parametric pump $A_2=0.635$~V, and spring constant modulation depth $\lambda=0.0465$. No double hysteresis can be seen.

\newpage

\section{Amplitude and phase responses of resonators~C and D as a function of $V_\mathrm{g}^\mathrm{dc}$}

\subsection{Responses of resonator~C}

\begin{figure}[H]
\centering
\includegraphics{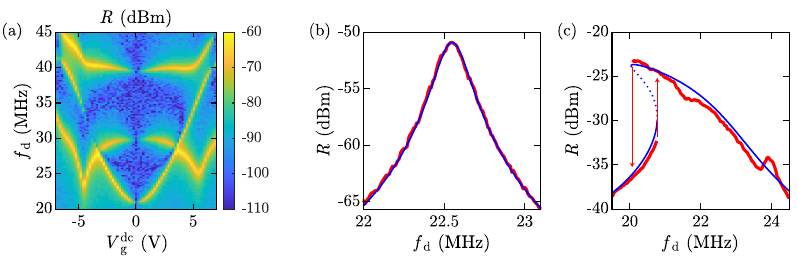}
\caption{Response of resonator~C to an external drive. (a) Amplitude response $R$ as a function of drive frequency $f_\mathrm{d}$ and gate voltage $V_\mathrm{g}^\mathrm{dc}$ in the absence of a parametric pump. The power of the external drive applied between the resonator and the gate is $P_\mathrm{d}=-37$~dBm. (b) Linear response at $V_\mathrm{g}^\mathrm{dc}=1.6$~V and $P_\mathrm{d}=-33$~dBm ($F_\mathrm{d}\simeq3\times10^{-11}$~N). The optical power incident on the resonator is $P_\mathrm{inc}\simeq14.1\times10^{-6}$~W. (c) Nonlinear response at $V_\mathrm{g}^\mathrm{dc}=1.6$~V and $P_\mathrm{d}=7$~dBm ($F_\mathrm{d}\simeq3\times10^{-9}$~N). $P_\mathrm{inc}\simeq14.7\times10^{-6}$~W. In (b) and (c), the red traces are measured data and the blue traces are fits of Eq.~(\ref{EQM}). $\mathrm{G}=2.5\times10^5$~V~W$^{-1}$, $\mathrm{T\simeq0.5}$, and $\vert\mathrm{d}r/\mathrm{d}z_\mathrm{s}\vert\simeq4.5\times10^{6}$~m$^{-1}$ in all panels.}\label{resonatorC1}
\end{figure}

Resonator~C has a similar shape as resonator~A in the main text. We estimate the effective mass of the fundamental mode to be $m_\mathrm{eff}\simeq m_\mathrm{membrane}/2\simeq7.2\times10^{-16}$~kg, where $m_\mathrm{membrane}$ is the mass of the suspended membrane estimated from atomic force microscopy images. The vibrations are measured at room temperature and in vacuum. Figure~\ref{resonatorC1}a shows the power $R$ of the signal at the ouptut of the photodetector for linear vibrations as a function of the gate voltage $V_\mathrm{g}^\mathrm{dc}$ and of the drive frequency $f_\mathrm{d}$, in the absence of a parametric pump. (See Supplementary Note~\ref{EQMsection} for the definition of $R$.) Figure \ref{resonatorC1}b shows $R(f_\mathrm{d})$ measured in the linear regime of vibrations. A fit of a Lorentzian lineshape allows us to estimate the damping rate $\tilde{\gamma}\simeq1.3\times10^6$~rad~s$^{-1}$. Figure \ref{resonatorC1}c shows $R(f_\mathrm{d})$ measured in the nonlinear regime of vibrations. A fit of the nonlinear response allows us to estimate the parameter of the nonlinear restoring force $\alpha\simeq-1.35\times10^{17}$~kg~m$^{-2}$~s$^{-2}$ and the parameter of the nonlinear dissipative force $\eta\simeq6\times10^8$~kg~m$^{-2}$~s$^{-1}$.

\begin{figure}[b]
\includegraphics{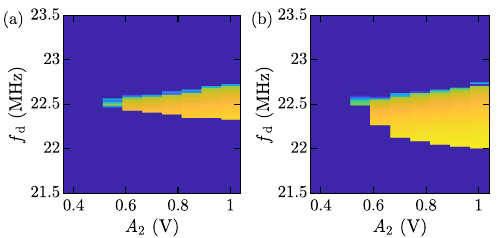}
\caption{Estimating the spring constant modulation depth $\lambda$ of resonator~C. A pump voltage $A_2\cos(2\times2\pi f_\mathrm{d}t)$ is applied between the resonator and the gate. The power of the signal at the output of the photodetector $R$ is shown as a function of $A_2$ upon increasing (a) and decreasing (b) $f_\mathrm{d}$. Gate voltage $V_\mathrm{g}^\mathrm{dc}=1.6$~V. Power scale: blue, $-80$~dBm; yellow, $-30$~dBm.}\label{resonatorC2}
\end{figure}

We now characterize the response of the resonator driven by a parametric pump.  Figure~\ref{resonatorC2} shows $R$ measured as a function of the peak voltage amplitude $A_2$ of the parametric pump and as a function of half the frequency of the pump $f_\mathrm{p}$, $f_\mathrm{d}=f_\mathrm{p}/2$. The gate voltage is $V_\mathrm{g}^\mathrm{dc}=1.6$~V, for which the resonant frequency of the fundamental mode is $\omega_0/2\pi\simeq22.6\times10^6$~Hz. $f_\mathrm{d}$ increases in Fig.~\ref{resonatorC2}a and decreases in Fig.~\ref{resonatorC2}b. The peak voltage amplitude above which a finite response $R$ is observed is the threshold pump voltage $A_{2,\mathrm{th}}\simeq0.5$~V. Hence the spring constant modulation depth $\lambda$ of resonator~C is $\lambda\simeq0.0366\,A_2$~V$^{-1}$ [Eq.~(\ref{lambda})].

\begin{figure}[h]
\includegraphics{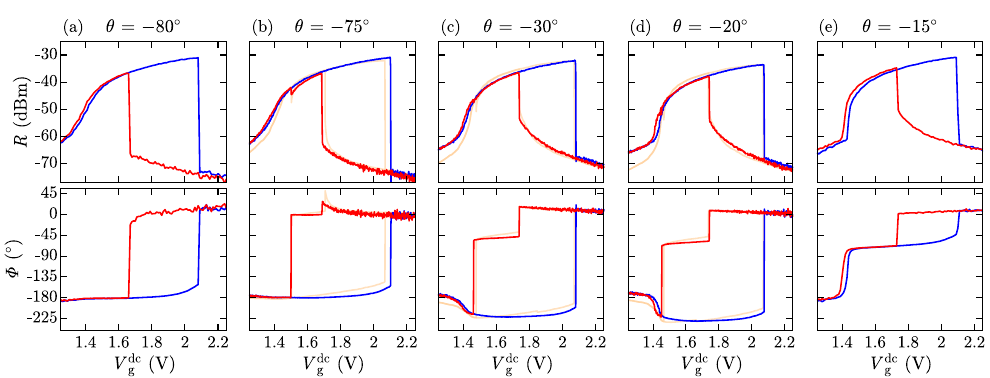}
\caption{Amplitude and phase responses, $R$ and $\Phi$, of the photodetector output signal of resonator~C as a function of $V_\mathrm{g}^\mathrm{dc}$ for various values of $\theta$, the phase difference between the external drive and the parametric pump. $A_1=1.4\times10^{-2}$~V, $A_2=1.26$~V, $\lambda\simeq0.046$, and $f_\mathrm{d}=22.6\times10^6$~Hz. The blue (red) traces are obtained upon increasing (decreasing) $V_\mathrm{g}^\mathrm{dc}$. From (a) to (e), $\theta=-80^\circ$, $-75^\circ$, $-30^\circ$, $-20^\circ$, and $-15^\circ$. The orange traces are fits of Eq.~(1) in the main text with the transduction factor $\kappa$ as a fit parameter.}\label{resonatorC5}
\end{figure}

Figure~\ref{resonatorC5} shows $R$ and the phase $\Phi$ of the signal at the output of the photodetector as a function of $V_\mathrm{g}^\mathrm{dc}$ for several values of $\theta$, the phase difference between the external drive and the parametric pump. $\Phi$ is measured at the input of the lock-in amplifier. We apply a compound excitation $\delta V_\mathrm{g}(t)=A_1\cos(2\pi f_\mathrm{d}t+\theta)+A_2\cos(2\times2\pi f_\mathrm{d}t)$ between the gate and the resonator, with $A_1=1.4\times10^{-2}$~V, $A_2=1.26$~V, and $f_\mathrm{d}=22.6\times10^6$~Hz. The blue (red) traces are obtained by increasing (decreasing) $V_\mathrm{g}^\mathrm{dc}$. The double hysteresis is not found for $\theta=-80^\circ$ but is observed for $\theta$ ranging from $-75^\circ$ to $-20^\circ$. It disappears again for $\theta=-15^\circ$.

\subsection{Responses of resonator~D}\label{responseofresonatorD}

\begin{figure}[h]
\includegraphics{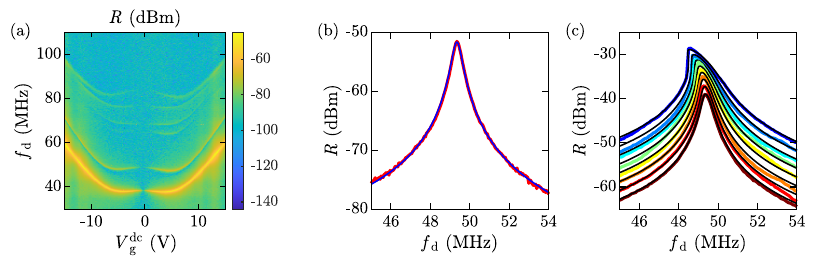}
\caption{Response of resonator~D to an external drive. (a) Amplitude response $R$ as a function of drive frequency $f_\mathrm{d}$ and gate voltage $V_\mathrm{g}^\mathrm{dc}$ in the absence of a parametric pump. The power of the external drive applied between the resonator and the gate is $P_\mathrm{d}=-37$~dBm. (b) Linear response at $V_\mathrm{g}^\mathrm{dc}=11$~V and $P_\mathrm{d}=-37$~dBm ($F_\mathrm{d}\simeq3\times10^{-11}$~N). The optical power incident on the resonator is $P_\mathrm{inc}\simeq6\times10^{-6}$~W. (c) Nonlinear response at $V_\mathrm{g}^\mathrm{dc}=11$~V for $P_\mathrm{d}$ ranging from $-27$~dBm ($F_\mathrm{d}\simeq10^{-10}$~N, bottom trace) to $-13$~dBm ($F_\mathrm{d}\simeq 5\times10^{-10}$~N, top trace). $P_\mathrm{inc}\simeq8.7\times10^{-6}$~W. In (b) and (c), the black traces are fits of Eq.~(\ref{EQM}) with the transduction factor $\kappa$ as a fit parameter.}\label{resonatorD1}
\end{figure}

Resonator~D is shaped as a drum with a diameter of 3~$\mu$m. We estimate the effective mass of the fundamental mode to be $m_\mathrm{eff}\simeq0.27\,m_\mathrm{membrane}\simeq6.3\times10^{-17}$~kg, where $m_\mathrm{membrane}$ is the mass of the suspended membrane estimated from atomic force microscopy images. The vibrations are measured at room temperature and in vacuum. Figure~\ref{resonatorD1}a shows the power $R$ of the signal at the ouptut of the photodetector for linear vibrations as a function of the gate voltage $V_\mathrm{g}^\mathrm{dc}$ and of the drive frequency $f_\mathrm{d}$, in the absence of a parametric pump. (See Supplementary Note~\ref{EQMsection} for the definition of $R$.) Figure \ref{resonatorD1}b shows $R(f_\mathrm{d})$ measured in the linear regime of vibrations. A fit of a Lorentzian lineshape allows us to estimate the damping rate $\tilde{\gamma}\simeq3.26\times10^6$~rad~s$^{-1}$. Figure \ref{resonatorD1}c shows $R(f_\mathrm{d})$ measured in the nonlinear regime of vibrations over a range of drive powers $P_\mathrm{d}$. A fit of the nonlinear responses allows us to estimate the parameter of the nonlinear restoring force $\alpha\simeq-1.35\times10^{16}$~kg~m$^{-2}$~s$^{-2}$ and the parameter of the nonlinear dissipative force $\eta\simeq2.7\times10^7$~kg~m$^{-2}$~s$^{-1}$.

\begin{figure}[h]
\includegraphics{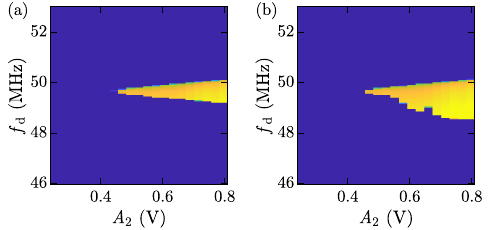}
\caption{Estimating the spring constant modulation depth $\lambda$ of resonator~D. A pump voltage $A_2\cos(2\times2\pi f_\mathrm{d}t)$ is applied between the resonator and the gate. The power of the signal at the output of the photodetector $R$ is shown as a function of $A_2$ upon increasing (a) and decreasing (b) $f_\mathrm{d}$. Gate voltage $V_\mathrm{g}^\mathrm{dc}=11$~V. Power scale: blue, $-80$~dBm; yellow, $-30$~dBm.}\label{resonatorD3}
\end{figure}

We now characterize the response of the resonator driven by a parametric pump.  Figure~\ref{resonatorD3} shows $R$ measured as a function of the peak voltage amplitude $A_2$ of the parametric pump and as a function of half the frequency of the pump $f_\mathrm{p}$, $f_\mathrm{d}=f_\mathrm{p}/2$. The gate voltage is $V_\mathrm{g}^\mathrm{dc}=11$~V, for which the resonant frequency of the fundamental mode is $\omega_0/2\pi\simeq49.4\times10^6$~Hz. $f_\mathrm{d}$ increases in Fig.~\ref{resonatorD3}a and decreases in Fig.~\ref{resonatorD3}b. The peak voltage amplitude above which a finite response $R$ is observed is the threshold pump voltage $A_{2,\mathrm{th}}\simeq0.4$~V. Hence the spring constant modulation depth $\lambda$ of resonator~D is $\lambda\simeq0.0525A_2$~V$^{-1}$ [Eq.~(\ref{lambda})].

\begin{figure}[h]
\includegraphics{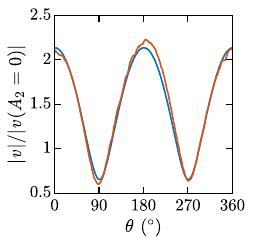}
\caption{Normalized amplitude of the output voltage of the photodetector as a function of $\theta$, the phase difference between the external drive and the parametric pump, below threshold. Resonator~D is measured. $A_1=4.2\times10^{-3}$~V and $A_2=0.21$~V. The red trace represents the measured data and the blue trace is a fit of Eq.~(\ref{paramgain}) with $\Delta\theta$ as a fit parameter.}\label{resonatorD4}
\end{figure}

We then characterize the parametric gain below the threshold for large amplitude oscillations. Figure~\ref{resonatorD4} shows $v=[2\langle V_\mathrm{PD}^2(t)\rangle]^{1/2}$, the peak amplitude of the voltage $V_\mathrm{PD}(t)$ at the output of the photodetector, normalized to its value without parametric pump ($A_2=0$), as a function of the phase difference $\theta$ between the external drive and the pump. The peak amplitude of the external drive is $A_1=4.2\times10^{-3}$~V and that of the pump is $A_2=0.21$~V. The following expression is fit to the data \cite{Lifshitz2009,Eichler2012}:
\begin{equation}
\frac{v(A_2)}{v(A_2=0)}=\left|\frac{\cos(\theta+\Delta\theta+\pi/4)}{1-A_2/A_{2,\mathrm{th}}}+\mathrm{i}\frac{\sin(\theta+\Delta\theta+\pi/4)}{1+A_2/A_{2,\mathrm{th}}}\right|\,,\label{paramgain}
\end{equation}
where $\Delta\theta$ is an extra phase offset between the driving force and the parametric excitation that arises from propagation delays. From the fit, we extract $\Delta\theta\simeq-46^\circ$.

\begin{figure}[h]
\includegraphics{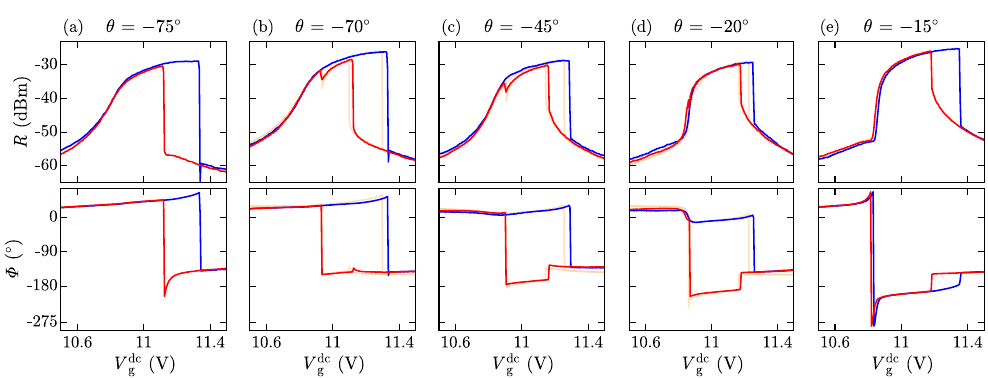}
\caption{Amplitude and phase responses, $R$ and $\Phi$, of the photodetector output signal of resonator~D as a function of $V_\mathrm{g}^\mathrm{dc}$ for various values of $\theta$, the phase difference between the external drive and the parametric pump. $A_1=7\times10^{-3}$~V, $A_2=0.64$~V, $\lambda\simeq0.0337$, and $f_\mathrm{d}=49.85\times10^6$~Hz. The blue (red) traces are obtained upon increasing (decreasing) $V_\mathrm{g}^\mathrm{dc}$. From (a) to (e), $\theta=-75^\circ$, $-70^\circ$, $-45^\circ$, $-20^\circ$, and $-15^\circ$. The orange traces are fits of Eq.~(1) in the main text with the transduction factor $\kappa$ as a fit parameter.}\label{resonatorD2}
\end{figure}

Figure~\ref{resonatorD2} shows $R$ and the phase $\Phi$ of the signal at the output of the photodetector as a function of $V_\mathrm{g}^\mathrm{dc}$ for several values of $\theta$. We apply a compound excitation $\delta V_\mathrm{g}(t)=A_1\cos(2\pi f_\mathrm{d}t+\theta)+A_2\cos(2\times2\pi f_\mathrm{d}t)$ between the gate and the resonator, with $A_1=7\times10^{-3}$~V, $A_2=0.64$~V, and $f_\mathrm{d}=49.85\times10^6$~Hz. The blue (red) traces are obtained by increasing (decreasing) $V_\mathrm{g}^\mathrm{dc}$. The double hysteresis is not found for $\theta=-75^\circ$ but is observed for $\theta$ ranging from $-70^\circ$ to $-20^\circ$. It disappears again for $\theta=-15^\circ$.

\newpage

\section{Calibrating the force noise applied to resonator~A}\label{calibration}

In the main text, we incoherently drive resonator~A with a force noise by applying a white Gaussian voltage noise of power spectral density $S_{uu}$ between the resonator and the gate. The power spectral density of the force noise is $S_{FF}=(C_\mathrm{g}^\prime V_\mathrm{g}^\mathrm{dc})^2S_{uu}$, where $C_\mathrm{g}^\prime$ is the derivative of the capacitance $C_\mathrm{g}$ between the resonator and the gate with respect to the static displacement of the resonator and $V_\mathrm{g}^\mathrm{dc}$ is the gate voltage. The effective thermal energy $k_\mathrm{B}T_\mathrm{eff}$ of the vibration mode is related to the variance of displacement $\langle\delta z^2\rangle$ through the equipartition theorem:
\begin{equation}
k_\mathrm{B}T_\mathrm{eff}=m_\mathrm{eff}\omega_0^2\langle\delta z^2\rangle\,,\label{Teff1}
\end{equation} 
with
\begin{equation}
\langle\delta z^2\rangle=S_\mathrm{FF}\int_0^\infty\mathrm{d}\omega\,\vert\chi(\omega)\vert^2\simeq S_\mathrm{FF}\frac{\pi Q(T_\mathrm{eff})}{2\omega_0^3}\,,\label{Teff2}
\end{equation} 
where $m_\mathrm{eff}$ is the effective mass, $\chi$ is the Fourier transform of the susceptibility in the frequency domain , $\omega_0$ is the angular resonant frequency, and $Q$ is the temperature-dependent quality factor of the fundamental vibration mode, respectively, and where the power spectral densities are single sided. Combining Eqs.~(\ref{Teff1}) and (\ref{Teff2}) yields:
\begin{equation}
k_\mathrm{B}T_\mathrm{eff}\simeq\frac{\pi m_\mathrm{eff}}{2\omega_0}\left(C_\mathrm{g}^\prime V_\mathrm{g}^\mathrm{dc}\right)^2S_{uu}Q(T_\mathrm{eff})\,.\label{Teff3}
\end{equation} 

\begin{figure}[h]
\includegraphics{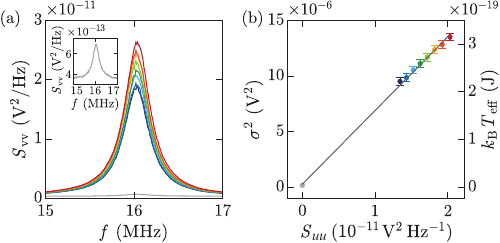}
\caption{Calibrating the force noise applied to resonator~A. (a) Power spectral density of voltage fluctuations $S_{vv}$ at the output of the photodetector as a function of spectral frequency $f$ for various power spectral densities $S_{uu}$ of voltage fluctuations applied between the resonator and the gate. $S_{vv}(f)$ at room temperature is shown as the grey trace in this panel and in the inset. (b) Left-hand-side ordinate axis: $\sigma^2$, the area under $S_{vv}(f)$. Right-hand-side ordinate axis: effective thermal energy $k_\mathrm{B}T_\mathrm{eff}$, with $k_\mathrm{B}$ the Boltzmann constant. The straight line is a linear fit. A spectrum in (a) and its corresponding dot in (b) have the same color.}\label{thermalnoise}
\end{figure}

We calibrate the force noise using the spectrum of thermal vibrations measured at room temperature (Figs.~\ref{thermalnoise}a, b),
\begin{equation}
T_\mathrm{eff}(S_{uu})=\frac{\sigma^2(S_{uu})}{\sigma^2_\mathrm{295K}}\times295\,\mathrm{K}\,, \label{Teffcal}
\end{equation}
where $\sigma^2$ is the area under the power spectral density of voltage fluctuations $S_{vv}$ at the output of the photodetector (recall that $S_{uu}$ is the power spectral density of voltage fluctuations applied to the gate) and $\sigma^2_\mathrm{295K}$ is $\sigma^2$ measured at room temperature. Note that 295~K is the temperature of the lab. We estimate that the incident laser beam heats up the resonator above this temperature only slightly. Indeed, the temperature increase $\Delta T$ caused by the laser beam can be estimated as \cite{Barton2012}:
\begin{equation}
\Delta T\simeq\frac{A\times P_\mathrm{inc}}{4\pi\mathrm{k}_\mathrm{th}h}\,,
\end{equation}
where $A$ is the absorbance of the membrane, $P_\mathrm{inc}$ is the intensity of the light incident on the resonator, $\mathrm{k}_\mathrm{th}$ is the thermal conductivity of the membrane, and $h$ is the thickness of the membrane. With $h\simeq4\times10^{-9}$~m (about 10 layers), $\mathrm{k}_\mathrm{th}\simeq1950$~W~m$^{-1}$~K$^{-1}$ for pyrolytic graphite (parallel to the layer planes and at room temperature) \cite{Ho1972}, and with the largest possible values $P_\mathrm{inc}=3\times10^{-5}$~W and $A=1$, we find $\Delta T\simeq0.3$~K.

\section{Activation of the transition rates between the two parametric phase states in resonator~A} \label{U}

\begin{figure}[h]
\includegraphics{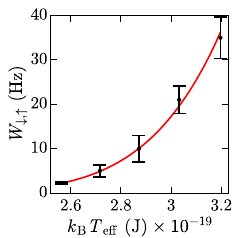}
\caption{Activation of the transition rates $W_{\uparrow,\downarrow}$ between the two parametric phase states of resonator~A. $W_{\uparrow,\downarrow}$ are measured as a function of the effective thermal energy $k_\mathrm{B}T_\mathrm{eff}$ with $A_1=0$ and $A_2=0.635$~V. The red trace is a fit of Eq.~(\ref{W}).}\label{activationfig}
\end{figure}

We consider resonator~A in the main text. Figure~\ref{activationfig} shows the transition rates $W_\uparrow$ out of state $\uparrow$ and $W_\downarrow$ out of state $\downarrow$, where $\uparrow$ ($\downarrow$) labels the phase state with $\Phi\simeq0$ ($\Phi\simeq-\pi$), as a function of the effective thermal energy $k_\mathrm{B}T_\mathrm{eff}$ induced by a white Gaussian force noise (Supplementary Note~\ref{calibration}). A parametric pump of amplitude $A_2=0.635$~V is applied in the absence of an external drive ($A_1=0$). We find that $W_\uparrow\simeq W_\downarrow$. We also find that $W_{\uparrow,\downarrow}$ follow an activation law \cite{Dykman1998,Lapidus1999,Marthaler2006,Kim2006,Chan2007,Dolleman2019,Han2024}:
\begin{equation}
W_{\uparrow,\downarrow}=C\exp\left(-U_{\uparrow,\downarrow}/k_\mathrm{B}T_\mathrm{eff}\right)\,,\label{W}
\end{equation}
with $C$ a constant and $U_\uparrow$ ($U_\downarrow$) the activation energy out of state $\uparrow$ ($\downarrow$). A fit of Eq.~(\ref{W}) to the data yields $U_\uparrow\simeq U_\downarrow\simeq3.6\times10^{-18}$~J.

\section{Measuring the logarithmic susceptibility of resonator~A}\label{logsuscept}

According to the theory in Ref.~\cite{Ryvkine2006}, the external drive $F_\mathrm{d}\cos(\omega_\mathrm{d}t+\theta)$ modifies the activation energies $U_{\uparrow,\downarrow}$ (see Supplementary Note~\ref{U}) as
\begin{equation}
U_{\uparrow,\downarrow}=\bar{U}+\sigma_{\uparrow,\downarrow}\chi F_\mathrm{d}\cos(\theta+\delta)\,,\label{logsusceptibility}
\end{equation}
where $\bar{U}$ is the activation energy in the absence of a symmetry breaking drive ($F_\mathrm{d}=0$), $\sigma_\uparrow=1$, $\sigma_\downarrow=-1$, and $\chi$ and $\delta$ are the magnitude and the phase of the logarithmic susceptibility of the resonator.

The logarithmic susceptibility can be extracted from the ratio of the transition rates $W_\downarrow/W_\uparrow$. Indeed, combining Eqs.~(\ref{W}) and (\ref{logsusceptibility}) yields
\begin{equation}
\mathrm{log}\left(\frac{W_\downarrow}{W_\uparrow}\right)=\frac{2\chi F_\mathrm{d}\cos(\theta+\delta)}{k_\mathrm{B}T_\mathrm{eff}}\,.\label{WupoverWdown}
\end{equation}

\begin{figure}[h]
\includegraphics{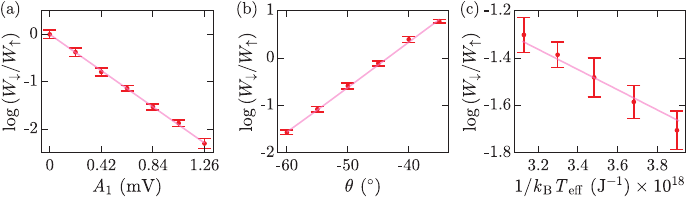}
\caption{Measuring the logarithmic susceptibility of resonator~A. (a) Natural logarithm of the ratio of the transition rates, $\mathrm{log}(W_\downarrow/W_\uparrow)$, as a function of $A_1$ with $\theta=-60^\circ$ and $k_\mathrm{B}T_\mathrm{eff}\simeq2.2\times10^{-19}$~J. The solid trace is a fit of the data to Eq.~(\ref{WupoverWdown}) with $\chi\cos(\theta+\delta)\simeq-3.1\times10^{-8}$~J~N$^{-1}$. (b) $\mathrm{log}(W_\downarrow/W_\uparrow)$ as a function of $\theta$ with $A_1=1.05\times10^{-3}$~V and $k_\mathrm{B}T_\mathrm{eff}\simeq2.6\times10^{-19}$~J. The solid trace is a fit of Eq.~(\ref{WupoverWdown}) with $\chi\simeq1.1\times10^{-7}$~J~N$^{-1}$ and $\delta\simeq-46.5^\circ$. (c) $\mathrm{log}(W_\downarrow/W_\uparrow)$ as a function of $1/k_\mathrm{B}T_\mathrm{eff}$ with $A_1=1.05\times10^{-3}$~V and $\theta=-60^\circ$. The solid trace is a fit of Eq.~(\ref{WupoverWdown}) with $\chi\cos(\theta+\delta)\simeq-3.2\times10^{-8}$~J~N$^{-1}$. $V_\mathrm{g}^\mathrm{dc}=2.4$~V, $\omega_0/2\pi\simeq\omega_\mathrm{d}/2\pi=16\times10^6$~Hz, and $A_2=0.635$~V in all panels.}\label{logarithmic}
\end{figure}

We consider resonator~A in the main text. Figures.~\ref{logarithmic}a-c show $\mathrm{log}(W_\downarrow/W_\uparrow)$ as a function of $A_1$ (with $\theta=-60^\circ$ and $k_\mathrm{B}T_\mathrm{eff}\simeq2.2\times10^{-19}$~J), $\theta$ (with $A_1=1.05\times10^{-3}$~V and $k_\mathrm{B}T_\mathrm{eff}\simeq2.6\times10^{-19}$~J), and $1/k_\mathrm{B}T_\mathrm{eff}$ ($A_1=1.05\times10^{-3}$~V and $\theta=-60^\circ$), all with $V_\mathrm{g}^\mathrm{dc}=2.4$~V, $A_2=0.635$~V, and $\omega_0/2\pi\simeq\omega_\mathrm{d}/2\pi=16\times10^6$~Hz. Fitting Eq.~(\ref{WupoverWdown}) to these data, we find similar values for the logarithmic susceptibility: $\chi\cos(\theta+\delta)\simeq-3.1\times10^{-8}$~J~N$^{-1}$ (Figs.~\ref{logarithmic}a, c), and $\chi\simeq1.1\times10^{-7}$~J~N$^{-1}$ and $\delta\simeq-46.5^\circ$ (Fig.~\ref{logarithmic}b).

\section{Occupation probabilities of the parametric phase states of resonator~C}

\begin{figure}[h]
\includegraphics{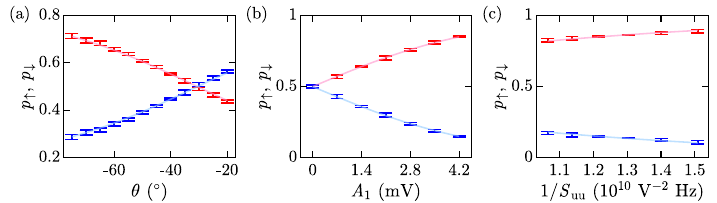}
\caption{Occupation probabilities $p_{\uparrow,\downarrow}$ of the parametric phase states of resonator~C in the presence of a white Gaussian force noise. (a) $p_{\uparrow,\downarrow}$ as a function of $\theta$ with $A_1=1.4\times10^{-3}$~V and $S_{uu}\simeq7.95\times10^{-11}$~V$^2$~Hz$^{-1}$. (b) $p_{\uparrow,\downarrow}$ as a function of $A_1$ with $\theta=-75^\circ$ and $S_{uu}\simeq1.326\times10^{-10}$~V$^2$~Hz$^{-1}$. (c) $p_{\uparrow,\downarrow}$ as a function of $1/S_{uu}$ with $A_1=2.8\times10^{-3}$~V and $\theta=-75^\circ$. Solid traces are fits of Eq.~(6) in the main text. Fit parameters are (a) $\chi\simeq8.66$~V$^2$~Hz$^{-1}$~N$^{-1}$, $\delta\simeq-59.5^\circ$; (b) $\chi\cos(\theta+\delta)\simeq-6.5$~V$^2$~Hz$^{-1}$~N$^{-1}$; (c) $\chi\cos(\theta+\delta)\simeq-6.02$~V$^2$~Hz$^{-1}$~N$^{-1}$. In all panels, $V_\mathrm{g}^\mathrm{dc}=1.6$~V, $A_2=1.26$~V, and $f_\mathrm{d}=22.6\times10^6$~Hz.}\label{resonatorC3}
\end{figure}

Figure~\ref{resonatorC3} shows the measured occupation probabilities $p_{\uparrow,\downarrow}$ of the two parametric phase states of resonator~C in the presence of a white Gaussian force noise. Figure~\ref{resonatorC3}a shows $p_{\uparrow,\downarrow}$ as a function of $\theta$, the phase difference between the external drive and the parametric pump. Figure~\ref{resonatorC3}b shows $p_{\uparrow,\downarrow}$ as a function of $A_1$, the amplitude of the external drive. Figure~\ref{resonatorC3}c shows $p_{\uparrow,\downarrow}$ as a function of $1/S_{uu}$, where $S_{uu}$ is the power spectral density of voltage fluctuations applied to the gate (the effective temperature of the vibrations $T_\mathrm{eff}$ was not measured for resonator~C; it is expected that $S_{uu}\propto T_\mathrm{eff}$, as measured in resonators A and D).

\begin{figure}[t]
\includegraphics{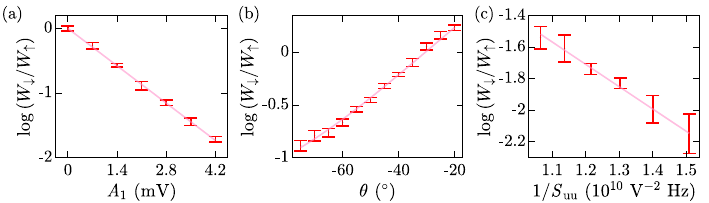}
\caption{Logarithmic susceptibility of resonator~C. (a) $\mathrm{log}[W_\downarrow/W_\uparrow]$ as a function of $A_1$ with $\theta=-75^\circ$ and $S_{uu}\simeq1.326\times10^{-10}$~V$^2$~Hz$^{-1}$. The solid trace is a fit of Eq.~(\ref{WupoverWdown}) with $\chi\cos(\theta+\delta)\simeq-6.5$~V$^2$~Hz$^{-1}$~N$^{-1}$. (b) $\mathrm{log}[W_\downarrow/W_\uparrow]$ as a function of $\theta$ with $A_1=1.4\times10^{-3}$~V and $S_{uu}\simeq7.95\times10^{-11}$~V$^2$~Hz$^{-1}$. The solid trace is a fit of Eq.~(\ref{WupoverWdown}) with $\chi\simeq8.66$~V$^2$~Hz$^{-1}$~N$^{-1}$ and $\delta\simeq-59.5^\circ$. (c) $\mathrm{log}[W_\downarrow/W_\uparrow]$ as a function of $1/S_{uu}$ with $A_1=2.8\times10^{-3}$~V and $\theta=-75^\circ$. The solid trace is a fit of Eq.~(\ref{WupoverWdown}) with $\chi\cos(\theta+\delta)\simeq-6.02$~V$^2$~Hz$^{-1}$~N$^{-1}$. In all panels, $V_\mathrm{g}^\mathrm{dc}=1.6$~V, $A_2=1.26$~V, and $f_\mathrm{d}=22.6\times10^6$~Hz.}\label{resonatorC4}
\end{figure}

Figure~\ref{resonatorC4} shows the natural logarithm of the ratio of the measured transition rates $W_{\uparrow,\downarrow}$ as a function of $A_1$, $\theta$, and $1/S_{uu}$. We employ $\mathrm{log}[W_\downarrow/W_\uparrow]$ to estimate the logarithmic susceptibility of the resonator (see Supplementary Note~\ref{logsuscept}).
\newpage

\section{Occupation probabilities of the parametric phase states of resonator~D}

\begin{figure}[h]
\includegraphics{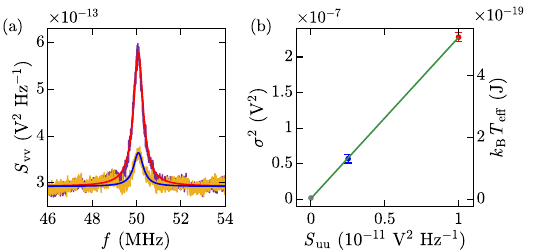}
\caption{Estimating the effective temperature of the fundamental vibration mode of resonator~D. (a) Power spectral density of voltage fluctuations $S_{vv}$ at the output of the photodetector as a function of the spectral frequency $f$ for $S_{uu}=2.5\times10^{-12}$~V$^2$~Hz$^{-1}$ (orange trace) and $S_{uu}=10^{-11}$~V$^2$~Hz$^{-1}$ (purple trace), where $S_{uu}$ is the power spectral density of voltage fluctuations applied to the gate. Blue and red curves are fits of the linear response. (b) Left-hand-side ordinate axis: area $\sigma^2$ under $S_{vv}(f)$. Right-hand-side ordinate axis: effective thermal energy $k_\mathrm{B}T_\mathrm{eff}$. The straight line is a linear fit. The blue dot and the red dots correspond to the red trace and the blue trace in (a). }\label{resonateurDSuu}
\end{figure}

We start by evaluating the effective temperature $T_\mathrm{eff}$ of the fundamental vibration mode of resonator~D in the presence of a white Gaussian force noise. With $S_{uu}=0$, where $S_{uu}$ is the power spectral density of voltage fluctuations applied to the gate, we do not resolve the thermal vibrations of the resonator at room temperature, hence we cannot calibrate $T_\mathrm{eff}$ as we do with resonator~A. (We have observed that room temperature thermal vibrations are visible in resonators whose resonant frequencies are $\simeq20$~MHz or below, while the amplitude response of resonator~D peaks near 50~MHz.) Instead, we quantify $T_\mathrm{eff}$ using the transduction factor $\kappa$ and the power spectral density of voltage fluctuations $S_{vv}$ at the output of the photodetector. The variance of displacement $\langle z^2\rangle$ reads:
\begin{equation}
\langle z^2\rangle=\frac{1}{\kappa^2}\int_{f_0-\Delta f}^{f_0+\Delta f}\mathrm{d}f\,S_{vv}(f)=\frac{\sigma^2}{\kappa^2}\,, \label{lavariancedeplacement}
\end{equation}  
where $f$ is the spectral frequency, $f_0$ is the resonant frequency, $2\Delta f$ is a wide enough frequency bandwidth encompassing the resonance in $S_{vv}(f)$, and $\sigma^2$ is the area under $S_{vv}(f)$. From the equipartition theorem, we obtain:
\begin{equation}
k_\mathrm{B}T_\mathrm{eff}=\frac{4\pi^2f_0^2m_\mathrm{eff}\sigma^2}{\kappa^2}\,,\label{TeffresonatorD}
\end{equation}
where $k_\mathrm{B}$ is the Boltzmann constant and $m_\mathrm{eff}$ is the effective mass of the vibration mode. The transduction factor $\kappa$ reads:
\begin{equation}
\kappa=P_\mathrm{inc}\times\mathrm{T}\times\mathrm{G}\times\left|\frac{\mathrm{d}r}{\mathrm{d}z_\mathrm{s}}\right|\,,\label{kapparesonatorD}
\end{equation}
where $P_\mathrm{inc}$ is the optical power incident on the resonator, $\mathrm{T}$ is the transmittance of the optical path from the resonator to the photodetector, $\mathrm{G}$ is the transimpedance gain of the photodetector, and $\mathrm{d}r/\mathrm{d}z_\mathrm{s}$ is the derivative with respect to the static displacement of the resonator $z_\mathrm{s}$ of the optical reflection coefficient $r$ at the surface of the resonator facing the light source. We measure $S_{vv}(f)$ for two values of $S_{uu}$ (Fig.~\ref{resonateurDSuu}a). With $P_\mathrm{inc}\simeq6\times10^{-6}$~W, $\mathrm{T}\simeq0.5$, $\mathrm{G}=2.5\times10^5$~V~W$^{-1}$, $\mathrm{d}r/\mathrm{d}z_\mathrm{s}\simeq4.5\times10^6$~m$^{-1}$ (based on our calculation), $f_0\simeq50\times10^6$~Hz, and $m_\mathrm{eff}\simeq6.3\times10^{-17}$~kg, we obtain an estimate of $k_\mathrm{B}T_\mathrm{eff}$ as a function of $S_{uu}$ (Fig.~\ref{resonateurDSuu}b).

\newpage

\begin{figure}[h]
\centering
\includegraphics{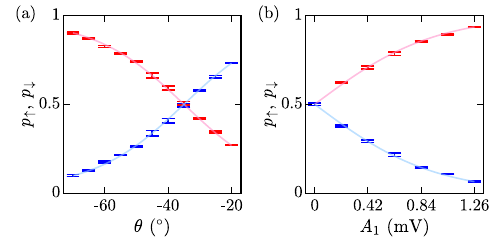}
\caption{Occupation probabilities $p_{\uparrow,\downarrow}$ of the parametric phase states of resonator~D in the presence of a white Gaussian force noise. (a) $p_{\uparrow,\downarrow}$ as a function of $\theta$ with $A_1=1.05\times10^{-3}$~V and $k_\mathrm{B}T_\mathrm{eff}\simeq5.2\times10^{-19}$~J. (b) $p_{\uparrow,\downarrow}$ as a function of $A_1$ with $\theta=-70^\circ$ and $k_\mathrm{B}T_\mathrm{eff}\simeq5.2\times10^{-19}$~J. The solid traces are fits of Eq.~(6) in the main text. Fit parameters are (a) $\chi\simeq1.52\times10^{-7}$~J~N$^{-1}$, $\delta\simeq-55^\circ$; (b) $\chi\cos(\theta+\delta)\simeq-8.75\times10^{-8}$~J~N$^{-1}$. In both panels, $V_\mathrm{g}^\mathrm{dc}=11$~V, $A_2=0.64$~V, and $f_\mathrm{d}=49.85\times10^7$~Hz.}\label{resonatorD6}
\end{figure}

Figure~\ref{resonatorD6} shows the measured occupation probabilities $p_{\uparrow,\downarrow}$ of the two parametric phase states of resonator~D in the presence of a white Gaussian force noise. Figure~\ref{resonatorD6}a shows $p_{\uparrow,\downarrow}$ as a function of $\theta$, the phase difference between the external drive and the parametric pump. Figure~\ref{resonatorD6}b shows $p_{\uparrow,\downarrow}$ as a function of $A_1$, the amplitude of the external drive.

\begin{thebibliography}{00}

\bibitem{Rugar1991}
D. Rugar and P. Gr\"{u}tter, Mechanical parametric amplification and thermomechanical noise squeezing, \textit{Phys. Rev. Lett.} \textbf{67}, no. 6, 699--702 (1991). \url{https://doi.org/10.1103/PhysRevLett.67.699}

\bibitem{Turner1998}
K. L. Turner, S. A. Miller, P. G. Hartwell, N. C. MacDonald, S. H. Strogatz, and S. G. Adams, Five parametric resonances in a microelectromechanical system, \textit{Nature} \textbf{396}, no. 6707, 149--152 (1998). \url{https://doi.org/10.1038/24122}

\bibitem{Carr2000}
D. W. Carr, S. Evoy, L. Sekaric, H. G. Craighead, and J. M. Parpia, Parametric amplification in a torsional microresonator, \textit{Appl. Phys. Lett.} \textbf{77}, no. 10, 1545--1547 (2000). \url{https://doi.org/10.1063/1.1308270}

\bibitem{Karabalin2010}
R. B. Karabalin, S. C. Masmanidis, and M. L. Roukes, Efficient parametric amplification in high and very high frequency piezoelectric nanoelectromechanical systems, \textit{Appl. Phys. Lett.} \textbf{97}, no. 18, 183101 (2010). \url{https://doi.org/10.1063/1.3505500}

\bibitem{Eichler2012}
A. Eichler, J. Chaste, J. Moser, and A. Bachtold, Parametric amplification and self-oscillation in a carbon nanotube resonator, \textit{Nano Lett.} \textbf{11}, no. 7, 2699--2703 (2011). \url{https://doi.org/10.1021/nl200950d}

\bibitem{Gieseler2012}
J. Gieseler, B. Deutsch, R. Quidant, and L. Novotny, Subkelvin parametric feedback cooling of a laser-trapped nanoparticle, \textit{Phys. Rev. Lett.} \textbf{109}, no. 10, 103603 (2012). \url{https://doi.org/10.1103/PhysRevLett.109.103603}

\bibitem{Lifshitz2009}
M. C. Lifshitz and R. Cross, Nonlinear Dynamics of Nanomechanical Resonators (2010). In Nonlinear Dynamics of Nanosystems (eds G. Radons, B. Rumpf and H.G. Schuster). \url{https://doi.org/10.1002/9783527629374.ch8}

\bibitem{Eichler2023}
A. Eichler and O. Zilberberg, ``Classical and Quantum Parametric Phenomena'', Oxford Graduate Texts (2023). \url{https://academic.oup.com/book/55246}

\bibitem{Mahboob2010b}
I. Mahboob, C. Froitier, and H. Yamaguchi, A symmetry-breaking electromechanical detector, \textit{Appl. Phys. Lett.} \textbf{96}, no. 21, 213103 (2010). \url{https://doi.org/10.1063/1.3429589}

\bibitem{Han2024}
C. Han, M. Wang, B. Zhang, M. I. Dykman, and H. B. Chan, Coupled parametric oscillators: From disorder-induced current to asymmetric Ising model, \textit{Phys. Rev. Research} \textbf{6}, no. 2, 023162 (2024). \url{https://doi.org/10.1103/PhysRevResearch.6.023162}

\bibitem{Dykman1998}
M. I. Dykman, C. M. Maloney, V. N. Smelyanskiy, and M. Silverstein, Fluctuational phase-flip transitions in parametrically driven oscillators, \textit{Phys. Rev. E} \textbf{ 57}, no. 5, 5202--5212 (1998). \url{https://doi.org/10.1103/PhysRevE.57.5202}

\bibitem{Lapidus1999}
L. J. Lapidus, D. Enzer, and G. Gabrielse, Stochastic phase switching of a parametrically driven electron in a penning trap, \textit{Phys. Rev. Lett.} \textbf{83}, no. 15, 899--902 (1999). \url{https://doi.org/10.1103/PhysRevLett.83.899}

\bibitem{Marthaler2006}
M. Marthaler and M. I. Dykman, Switching via quantum activation: A parametrically modulated oscillator, \textit{Phys. Rev. A} \textbf{73}, no. 4, 042108 (2006). \url{https://doi.org/10.1103/PhysRevA.73.042108}

\bibitem{Kim2006}
K. Kim, M. S. Heo, K. H. Lee, K. Jang, H. R. Noh, D. Kim, and W. Jhe, Spontaneous symmetry breaking of population in a nonadiabatically driven atomic trap: An Ising-class phase transition, \textit{Phys. Rev. Lett.} \textbf{96}, no. 15, 150601 (2006). \url{https://doi.org/10.1103/PhysRevLett.96.150601}

\bibitem{Chan2007}
H. B. Chan and C. Stambaugh, Activation barrier scaling and crossover for noise-induced switching in micromechanical parametric oscillators, \textit{Phys. Rev. Lett.} \textbf{99}, no. 6, 060601 (2007). \url{https://doi.org/10.1103/PhysRevLett.99.060601}

\bibitem{Goto1959}
E. Goto, The Parametron, a Digital Computing Element Which Utilizes Parametric Oscillation, in \textit{Proceedings of the IRE} \textbf{47}, no. 8, 1304--1316 (1959). \url{https://www.doi.org/10.1109/JRPROC.1959.287195}

\bibitem{Mahboob2008}
I. Mahboob and H. Yamaguchi, Bit storage and bit flip operations in an electromechanical oscillator, \textit{Nat. Nanotechnol.} \textbf{3}, no. 5, 275--279 (2008). \url{https://doi.org/10.1038/nnano.2008.84}

\bibitem{Wang2013}
Z. Wang, A. Marandi, K. Wen, R. L. Byer, and Y. Yamamoto, Coherent Ising machine based on degenerate optical parametric oscillators, \textit{Phys. Rev. A} \textbf{88}, no. 6, 063853 (2013). \url{https://doi.org/10.1103/PhysRevA.88.063853}

\bibitem{McMahon2016}
P. L. McMahon, A. Marandi, Y. Haribara, R. Hamerly, C. Langrock, S. Tamate, T. Inagaki, H. Takesue, S. Utsunomiya, K. Aihara, R. L. Byer, M. M. Fejer, H. Mabuchi, and Y. Yamamoto, A fully programmable 100-spin coherent Ising machine with all-to-all connections, \textit{Science} \textbf{354}, no. 6312, 614--617 (2016). \url{https://doi.org/10.1126/science.aah5178}

\bibitem{Heugel2022}
T. L. Heugel, O. Zilberberg, C.Marty, R. Chitra, and A. Eichler, Ising machines with strong bilinear coupling, \textit{Phys. Rev. Res.} \textbf{4}, no. 1, 013149 (2022). \url{https://doi.org/10.1103/PhysRevResearch.4.013149}

\bibitem{Alvarez2024}
P. \'{A}lvarez, D. Pittilini, F. Miserocchi, S. Raamamurthy, G. Margiani, O. Ameye, J. del Pino, O. Zilberberg, and A. Eichler, Biased Ising Model Using Two Coupled Kerr Parametric Oscillators with External Force, \textit{Phys. Rev. Lett.} \textbf{132}, no. 20, 207401 (2024). \url{https://doi.org/10.1103/PhysRevLett.132.207401}

\bibitem{Leuch2016}
A. Leuch, L. Papariello, O. Zilberberg, C. L. Degen, R. Chitra, and A. Eichler, Parametric symmetry breaking in a nonlinear resonator, \textit{Phys. Rev. Lett.} \textbf{117}, no. 21, 214101 (2016). \url{https://doi.org/10.1103/PhysRevLett.117.214101}

\bibitem{Ryvkine2006}
D. Ryvkine and M. I. Dykman, Resonant symmetry lifting in a parametrically modulated oscillator, \textit{Phys. Rev. E} \textbf{74}, no. 6, 061118 (2006). \url{https://doi.org/10.1103/PhysRevE.74.061118}

\bibitem{Frimmer2019}
M. Frimmer, T. L. Heugel, \v{Z}. Nosan, F. Tebbenjohanns, D. H\"{a}lg, A. Akin, C. L. Degen, L. Novotny, R. Chitra, O. Zilberberg, and A. Eichler, Rapid flipping of parametric phase states, \textit{Phys. Rev. Lett.} \textbf{123}, no. 25, 254102 (2019). \url{https://doi.org/10.1103/PhysRevLett.123.254102}

\bibitem{Nosan2019}
\v{Z}. Nosan, P. M\"{a}rki, N. Hauff, C. Knaut, and A. Eichler, Gate-controlled phase switching in a parametron, \textit{Phys. Rev. E} \textbf{99}, no. 6, 062205 (2019). \url{https://doi.org/10.1103/PhysRevE.99.062205}

\bibitem{Papariello2016}
L. Papariello, O. Zilberberg, A. Eichler, and R. Chitra, Ultrasensitive hysteretic force sensing with parametric nonlinear oscillators, \textit{Phys. Rev. E} \textbf{94}, no. 2, 022201 (2016). \url{https://doi.org/10.1103/PhysRevE.94.022201}

\bibitem{Supplemental_Material}
See Supplementary Information for the response of the resonator to the external drive, to the parametric pump, and to the compound excitation; the method to estimate the parameters of the equation of motion of resonator~A; the method to estimate the spring constant modulation depth $\lambda$ of resonator~A; the measurements of stochastic frequency jumps upon changing $f_\mathrm{d}$ in resonator~B; the amplitude and the phase responses of resonator~A as a function of $V_\mathrm{g}^\mathrm{dc}$ at $\theta=-30^\circ$; the amplitude and the phase responses of resonators~C and D as a function of $V_\mathrm{g}^\mathrm{dc}$; the calibration of the force noise applied to resonator~A; a study of the activation of the transition rates between the two parametric phase states in resonator~A; the measurements of the logarithmic susceptibility of resonator~A; the measurements of the occupation probabilities of the parametric phase states of resonator~C; and the measurements of the occupation probabilities of the parametric phase states of resonator~D. The Supplementary Information also contains Ref.~\cite{Ho1972}.

\bibitem{Ho1972}
C. Y. Ho, R. W. Powell, and P. E. Liley, Thermal Conductivity of the Elements, \textit{J. Phys. Chem. Ref. Data} \textbf{1}, 279--421 (1972). \url{https://doi.org/10.1063/1.3253100}

\bibitem{Lemme2020}
M. C. Lemme, S. Wagner, K. Lee, X. Fan, G. J. Verbiest, S. Wittmann, S. Lukas, R. J. Dolleman, F. Niklaus, H. S. J. van der Zant, G. S. Duesberg, and P. G. Steeneken, Nanoelectromechanical sensors based on suspended 2D materials, \textit{Research} \textbf{2020}, 1--25, Art. no. 8748602 (2020).

\bibitem{Xu2022}
B. Xu, P. Zhang, J. Zhu, Z. Liu, A. Eichler, X.-Q. Zheng, J. Lee, A. Dash, S. More, S. Wu, Y. Wang, H. Jia, A. Naik, A. Bachtold, R. Yang, P. X.-L. Feng, and Z. Wang, Nanomechanical Resonators: Toward Atomic Scale. \textit{ACS Nano} \textbf{16}, no. 10, 15545--15585 (2022). \url{https://doi.org/10.1021/acsnano.2c01673}

\bibitem{Sazonova2004}
V. Sazonova, Y. Yaish, H. \"{U}st\"{u}nel, D. Roundy, T. A. Arias, and P. L. McEuen, A tunable carbon nanotube electromechanical oscillator, \textit{Nature} \textbf{431}, no. 7006, 284--287 (2004). \url{https://doi.org/10.1038/nature02905}

\bibitem{Unterreithmeier2009}
Q. Unterreithmeier, E. Weig, and J. Kotthaus, Universal transduction scheme for nanomechanical systems based on dielectric forces, \textit{Nature} \textbf{458}, no. 7241, 1001--1004 (2009). \url{https://doi.org/10.1038/nature07932}

\bibitem{Mathew2016}
J. P. Mathew, R. N. Patel, A. Borah, R. Vijay, and M. M. Deshmukh, Dynamical strong coupling and parametric amplification of mechanical modes of graphene drums, \textit{Nat. Nanotech.} \textbf{11}, no. 9, 747--751 (2016). \url{https://doi.org/10.1038/nnano.2016.94}

\bibitem{Su2021}
Z.-J. Su, Y. Ying, X.-X. Song, Z.-Z. Zhang, Q.-H. Zhang, G. Cao, H.-O. Li, G.-C. Guo, and G.-P. Guo, Tunable parametric amplification of a graphene nanomechanical resonator in the nonlinear regime, \textit{Nanotechnology} \textbf{32}, no. 15, 155203 (2021). \url{https://doi.org/10.1088/1361-6528/abc9ea}

\bibitem{Aldridge2005}
J. S. Aldridge and A. N. Cleland, Noise-Enabled Precision Measurements of a Duffing Nanomechanical Resonator, \textit{Phys. Rev. Lett.} \textbf{94}, no. 15, 156403 (2005). \url{https://doi.org/10.1103/PhysRevLett.94.156403}

\bibitem{Kozinsky2006}
I. Kozinsky, H. W. Ch. Postma, I. Bargatin, and M. L. Roukes, Tuning nonlinearity, dynamic range, and frequency of nanomechanical resonators, \textit{Appl. Phys. Lett.} \textbf{88}, no. 25, 253101 (2006). \url{https://doi.org/10.1063/1.2209211}

\bibitem{Chen2009}
C. Chen, S. Rosenblatt, K. I. Bolotin, W. Kalb, P. Kim, I. Kymissis, H. L. Stormer, T. F. Heinz, and J. Hone, Performance of monolayer graphene nanomechanical resonators with electrical readout, \textit{Nat. Nanotech.} \textbf{4}, no. 12, 861--867 (2009). \url{https://doi.org/10.1038/nnano.2009.267}

\bibitem{Unterreithmeier2010}
Q. P. Unterreithmeier, T. Faust, and J. P. Kotthaus, Nonlinear switching dynamics in a nanomechanical resonator, \textit{Phys. Rev. B} \textbf{81}, no. 24, 241405R (2010). \url{https://doi.org/10.1103/PhysRevB.81.241405}

\bibitem{Eichler2011}
A. Eichler, J. Moser, J. Chaste, M. Zdrojek, I. Wilson-Rae, and A. Bachtold, Nonlinear damping in mechanical resonators made from carbon nanotubes and graphene, \textit{Nat. Nanotech.} \textbf{6}, no. 6, 339--342 (2011). \url{https://doi.org/10.1038/nnano.2011.71}

\bibitem{Dykman1975}
M. I. Dykman and M. A. Krivoglaz, Spectral distribution of nonlinear oscillators with nonlinear friction due to a medium, \textit{Phys. Status Solidi B} \textbf{68}, no. 1, 111--123 (1975). \url{https://doi.org/10.1002/pssb.2220680109}

\bibitem{Zaitsev2012}
S. Zaitsev, O. Shtempluck, E. Buks, and O. Gottlieb, Nonlinear damping in a micromechanical oscillator, \textit{Nonlinear Dyn.} \textbf{67}, no. 1, 859--883 (2012). \url{https://doi.org/10.1007/s11071-011-0031-5}

\bibitem{vanderZande2010}
A. M. van der Zande, R. A. Barton, J. S. Alden, C. S. Ruiz-Vargas, W. S. Whitney, P. H. Q. Pham, J. Park, J. M. Parpia, H. G. Craighead, and P. L. McEuen, Large-Scale Arrays of Single-Layer Graphene Resonators, \textit{Nano. Lett.} \textbf{10}, no. 12, 4869--4873 (2010). \url{https://doi.org/10.1021/nl102713c}

\bibitem{Bunch2007}
J. S. Bunch, A. M. V. D. Zande, S. S. Verbridge, I. W. Frank, D. M. Tanenbaum, J. M. Parpia, H. G. Craighead, and P. L. McEuen, Electromechanical resonators from graphene sheets, \textit{Science} \textbf{315}, no. 5811, 490--493 (2007). \url{https://doi.org/10.1126/science.1136836}

\bibitem{Barton2012}
R. A. Barton, I. R. Storch, V. P. Adiga, R. Sakakibara, B. R. Cipriany, B. Ilic, S. P. Wang, P. Ong, P. L. McEuen, J. M. Parpia, and H. G. Craighead, Photothermal self-oscillation and laser cooling of graphene optomechanical systems, \textit{Nano Lett.} \textbf{12}, no. 9, 4681--4686 (2012). \url{https://doi.org/10.1021/nl302036x}

\bibitem{Davidovikj2016}
D. Davidovikj, J. J. Slim, S. J. Cartamil-Bueno, H. S. J. van der Zant, P. G. Steeneken, and W. J. Venstra, Visualizing the motion of graphene nanodrums, \textit{Nano Lett.} \textbf{16}, no. 4, 2768--2773 (2016). \url{https://doi.org/10.1021/acs.nanolett.6b00477}

\bibitem{Lu2021}
H. Lu, C. Yang, Y. Tian, J. Lu, F. Xu, C. Zhang, F. Chen, Y. Ying, K. G. Sch\"{a}dler, C. Wang, F. H. L. Koppens, A. Reserbat-Plantey, and J. Moser, Imaging vibrations of electromechanical few layer graphene resonators with a moving vacuum enclosure, \textit{Precis. Eng.} \textbf{72}, 769--776 (2021). \url{https://doi.org/10.1016/j.precisioneng.2021.06.012}

\bibitem{Dolleman2019}
R. J. Dolleman, P. Belardinelli, S. Houri, H. S. J. van der Zant, F. Alijani, and P. G. Steeneken, High-Frequency Stochastic Switching of Graphene Resonators Near Room Temperature, \textit{Nano Lett.} \textbf{19}, 1282--1288 (2019). \url{http://dx.doi.org/10.1021/acs.nanolett.8b04862}

\bibitem{Okamoto2013}
H. Okamoto, A. Gourgout, C.-Y. Chang, K. Onomitsu, I. Mahboob, E. Yi Chang, and H. Yamaguchi, Coherent phonon manipulation in coupled mechanical resonators, \textit{Nat. Phys.} \textbf{9}, no. 8, 480--484 (2013). \url{https://doi.org/10.1038/nphys2665}

\bibitem{Faust2013}
T. Faust, J. Rieger, M. J. Seitner, J. P. Kotthaus, and E. M. Weig, Coherent control of a classical nanomechanical two-level system, \textit{Nat. Phys.} \textbf{9}, no. 8, 485--488 (2013). \url{https://doi.org/10.1038/nphys2666}

\bibitem{ZZZhang2020}
Z.-Z. Zhang, X.-X. Song, G. Luo, Z.-J. Su, K.-L. Wang, G. Cao, H.-O. Li, M. Xiao, G.-C. Guo, L. Tian, G.-W. Deng, and G.-P. Guo, Coherent phonon dynamics in spatially separated graphene mechanical resonators, \textit{Proc. Natl. Acad. Sci. U.S.A.} \textbf{117}, no. 11, 5582--5587 (2020). \url{https://doi.org/10.1073/pnas.1916978117}

\end{thebibliography}
\end{document}